\RequirePackage{fix-cm}
\documentclass[twocolumn]{svjour3}          
\smartqed  

\usepackage{hyperref}
\usepackage{microtype}
\usepackage[american]{babel}
\usepackage[numbers]{natbib}
\usepackage{amsmath,amssymb,amsfonts}
\usepackage{algorithmic}
\usepackage{graphicx}
\usepackage{textcomp}
\usepackage{xcolor}
\usepackage{algorithm}
\usepackage{float}
\usepackage{setspace}
\usepackage{subfigure}
\usepackage{verbatim}
\usepackage[misc]{ifsym}
\usepackage{enumitem}
\usepackage{url}
\allowdisplaybreaks[4]

\newtheorem{Def}{Definition}
\newtheorem{thm}{Theorem}
\newtheorem{exm}{Example}
\newtheorem{Rem}{Remark}
\newtheorem{Prom}{Problem}

\newtheorem{Coro}{Corollary}
\newtheorem{proofnoof}{Proof}

\journalname{The VLDB Journal}
\begin{document}

\title{Pricing Private Data with Personalized Differential Privacy and Partial Arbitrage Freeness
}


\author{Shuyuan Zheng         \and
        Yang Cao \and
        Masatoshi Yoshikawa
}


\institute{
This paper is extended from the authors' previous work \cite{zheng2020money}.\\
    The authors are with the Department of Social Informatics, Kyoto University, Kyoto, Japan, 606-8501. \\
              \email{\{caryzheng@db.soc., yang@, yoshikawa@\}i.kyoto-u.ac.jp} \\
              Corresponding author: Yang Cao
}

\date{Received: date / Accepted: date}

\maketitle

\begin{abstract}
There is a growing trend regarding perceiving personal data as a commodity. 
Existing studies have built frameworks and theories about how to determine an arbitrage-free price of a given query according to the privacy loss quantified by differential privacy.
However, those studies have assumed that data buyers can purchase query answers with the arbitrary privacy loss of data owners, which may not be valid under strict privacy regulations and data owners’ increasing privacy concerns.
In this paper, we study how to empower data owners to control privacy loss in data trading.
First, we propose a framework for trading personal data that enables data owners to bound their personalized privacy losses.
Second, since bounded privacy losses indicate bounded utilities of query answers, we propose a reasonable relaxation of arbitrage freeness named \textit{partial arbitrage freeness}, i.e., the guarantee of arbitrage-free pricing only for a limited range of utilities, which provides more possibilities for our market design. 
Third, to avoid arbitrage, we propose a general method for ensuring \textit{arbitrage freeness} under personalized differential privacy.
Fourth, to fully utilize data owners' personalized privacy loss bounds, we propose privacy budget allocation techniques to allocate privacy losses for queries under arbitrage freeness. 
Finally, we conduct experiments to verify the effectiveness of our proposed trading protocols.
\end{abstract}

\section*{Declarations}

\textbf{Funding:} This work is partially supported by JSPS KAKENHI Grant No. 17H06099, 18H04093, 19K20269, 21J23090 and 21K19767.
\\
\textbf{Conflicts of interest/Competing interests:} The authors have no relevant financial or non-financial interests to disclose.
\\
\textbf{Availability of data and material:}
The datasets generated during and/or analysed during the current study are available in the dataset directory, \url{https://github.com/teijyogen/DataMarket-PDP-PAF}.
\\
\textbf{Code availability:}
The code for running the experiments present in this paper is available in the experiment directory, \url{https://github.com/teijyogen/DataMarket-PDP-PAF}.

\section{Introduction}

\subsection{Background}
Personal data, the new oil of the digital era, is extraordinarily valuable to individuals and organizations with regard to discovering knowledge and improving products or services. 
While a single individual's personal information is worth nothing in practice, its aggregate, i.e., personal data, can be worth billions \cite{Brustein_2012}. 
However, since personal data may also release sensitive information that can be used to identify individuals for criminal purposes, individuals also deserve appropriate compensation due to their potential \textit{privacy loss}. 
In particular, a study found that compensation, especially monetary compensation, reduces people's expectations for privacy protection \cite{Gabisch2014The}, which implies that some individuals would like to provide privacy in exchange for money. 
In fact, there is a growing trend towards \textit{personal data trading} perceiving personal data as a commodity, which meets the demand of both data buyers (i.e., who want to utilize personal data) and data owners (i.e., who generate the data). 
Some startup companies, such as Airbloc\footnote{https://airbloc.org/}, Datum\footnote{https://datum.org/}, Datacoup\footnote{http://datacoup.com/} and CitizenMe\footnote{https://citizenme.com/}, consider a personal data trading platform that connects data owners and data buyers directly as a new business model. 

\begin{figure}[t]
\centerline{\includegraphics[scale=0.2]{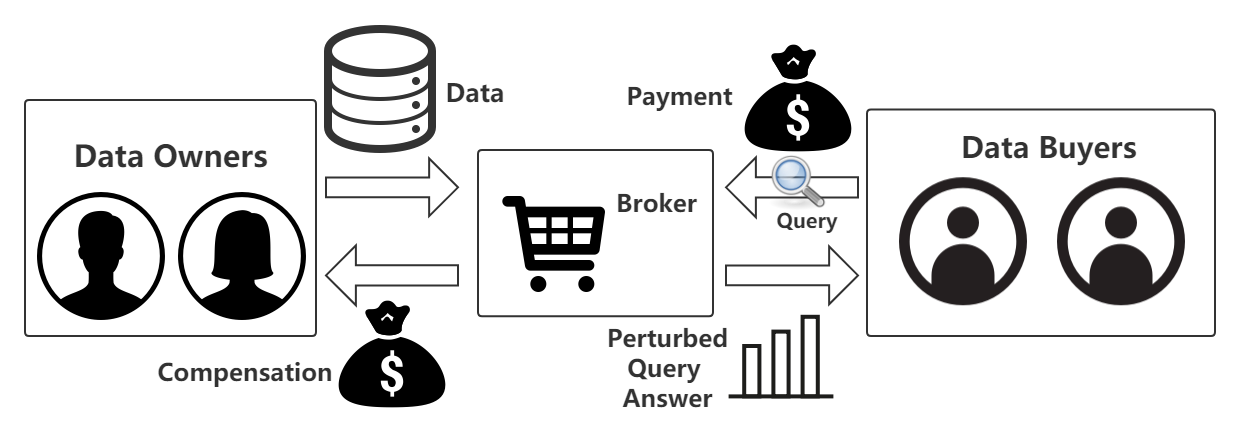}}
\caption{Query-based data trading.}
\label{fig:data_marketplace}
\end{figure}

In this work, we advocate for controllable privacy loss in data trading. 
However, the existing data marketplace frameworks do not give data owners the power to control their privacy loss.
Several studies \cite{ghosh2011selling, ghosh2015selling,li2013theory,li2014theory,jung2019privacy,chen2019towards,liu2021dealer} in the literature have investigated privacy-preserving query-based data trading, as shown in Figure \ref{fig:data_marketplace}.
There are three parties in the data marketplace, namely, \textit{data owners}, \textit{data buyers}, and a \textit{broker}. 
Data owners contribute their personal data and obtain monetary compensation in return. 
Data buyers request queries concerning the data and pay for perturbed query answers where some random noise is injected for privacy protection. 
The broker acts as a trustworthy intermediary between data owners and data buyers who is in charge of computing perturbed query answers, setting \textit{query prices}, and compensating data owners. 
A major challenge in this line of works is how to price queries. 
A seminal work by Li et al. \cite{li2014theory} assigned prices to perturbed queries according to their utility. The authors designed pricing functions by making the connection between the privacy loss and the utility; then, they formulated an important property of the pricing functions, namely, \textit{arbitrage freeness}, which refers to the consistency of query prices. 
Intuitively, a data buyer should not obtain an answer to a query by deriving this answer from a less expensive set of queries. 


\subsection{Gaps and Challenges}

There are two desiderata for a personal data market that have not been well studied in the literature.
First, data owners should be able to specify an \textit{upper bound} of privacy loss given the high sensitivity of personal data such as human trajectories \cite{de2013unique}. 
Second, data owners might have quite different privacy preferences, which calls for a personalized privacy loss bound.
A recent survey \cite{nget2017balance} revealed that most data owners value privacy more than financial compensation when monetizing their data, which motivates this work to guarantee data owners with personalized and controllable privacy loss in data trading. 
We identify the following four challenges to fulfill this goal.

First, there is no framework in the existing studies for trading data under bounded privacy loss. 
Bounding each data owner's privacy loss indicates the limited supply of the utility of the query answers since a lower bound of noise adding is needed to ensure the required privacy. 
This leads to a new and challenging problem, i.e., how to trade query answers when their utilities are bounded.

Second, personalized privacy losses make designing an arbitrage-free pricing function more difficult. 
As the price depends on the query answer's utility, while the cost of the answer (i.e., the compensations to data owners) is decided by the privacy losses of data owners, we design a pricing function based on a one-to-one connection between the utility and the privacy losses so that the price can cover the cost. Under uniform privacy losses, such a one-to-one connection is simple; however, under personalized privacy losses, the connection might be too complex to express in analytic form, which makes it extremely difficult to study its properties and guarantee the arbitrage freeness of the corresponding pricing function. Although we can leverage \textit{personalized differential privacy} (PDP) \cite{jorgensen2015conservative} to realize personalized privacy losses, no existing work has studied arbitrage-free pricing under PDP techniques, which is a nontrivial challenge. 

Third, it is difficult to make full use of data owners' personalized privacy loss bounds. 
Under uniform privacy loss bounds, we can simply consume uniform privacy losses to perturb query answers; finally, the privacy loss bounds can be exhausted.
However, we cannot make full use of personalized privacy loss bounds by uniform privacy losses.
Moreover, if we allocate personalized privacy losses to perturb query answers, the prices of those perturbed query answers might not be arbitrage free. 
Therefore, we need an intelligent tool to allocate privacy losses for query perturbation, by which data owners' revenues can be maximized under the constraint of arbitrage freeness.


\subsection{Contribution}

In this paper, for the first time, we study how to design a personal data marketplace that empowers data owners to control privacy loss.
Our main contributions are summarized as follows.

First, we design a framework and a reasonable relaxation of arbitrage freeness for trading personal data under bounded privacy losses. To match supply (i.e., bounded privacy losses from data owners) and demand (i.e., utility requirements from data buyers) in the data marketplace, our framework works with three phases in the trading protocol: \textit{Offer}, \textit{Quote}, and \textit{Delivery}. Our modularized framework provides a principled way to instantiate its trading protocol so that existing techniques can be easily adopted to power a data marketplace. 
Furthermore, as the utilities are also bounded under bounded privacy losses, we propose a practical relaxation of arbitrage freeness, i.e., \textit{partial arbitrage freeness}. As an important but rigorous constraint, arbitrage freeness always compresses the space of market design. For example, to achieve arbitrage freeness, the marketplace proposed in \cite{li2014theory} requires that each data owner's compensation should be a \textit{subadditive} function of privacy loss and that the perturbation mechanism should be the Laplace mechanism. However, in some cases where the utilities of query answers are bounded, we only need to guarantee arbitrage freeness for those queries with a certain range of utilities, i.e., to guarantee partial arbitrage freeness. By the introduction of partial arbitrage freeness, there can be more possibilities in designing a data marketplace, including more types of compensation functions and more perturbation mechanisms being provided for selection, while the marketplace is still robust to arbitrage behaviors.

Second, we propose a general method for ensuring arbitrage freeness or partial arbitrage freeness, which is significantly challenging under personalized privacy losses, as explained above.
Essentially, pricing a query requires the broker to derive data owners' privacy losses given a required utility.
Therefore, we design a pricing function based on the inverse function of the utility function (i.e., a mapping from the privacy losses to the utility). Then, we propose general theorems for arbitrage freeness and partial arbitrage freeness (Theorems \ref{theorem:arbitrage free}, \ref{thm:partial_ab} and \ref{thm:partial_ab_sup}), which present that if the utility function satisfies certain properties, it results in an arbitrage-free or partially arbitrage-free pricing function. Based on those theorems, we transform the problem of designing an arbitrage-free
or partially arbitrage-free pricing function into guaranteeing the certain properties of the utility function.

Third, to make full use of data owners' personalized privacy loss bounds, we propose online privacy budget allocation techniques, \textit{pattern} and 
\textit{PatternExchange}, to allocate privacy losses in an arbitrage-free or partially arbitrage-free manner. Concretely, we allocate privacy losses
by a fixed pattern (i.e., the set of ratio relationships between privacy losses) so that the pattern controls the properties of the utility function. In this way, by choosing an appropriate pattern for privacy loss allocation, we can indirectly ensure the arbitrage freeness or partial arbitrage freeness of the pricing function. Then, to promote the use of data owners' privacy loss bounds, we propose the PatternExchange algorithm to enable a dynamically changing pattern. Basically, a change in the pattern may cause a violation of arbitrage-free pricing; however, PatternExchange can flexibly change the pattern for each query and avoid such a violation.

The rest of this paper is organized as follows. In Sec. \ref{sec:prom}, we introduce some basic settings of our marketplace. Then, we propose our trading framework in Sec. \ref{sec:framework}, and provide some arbitrage-free and partially arbitrage-free trading protocols in Sec. \ref{sec:ab_protocols} and \ref{sec:pab_protocols}. We also present some experimental results on those trading protocols in Sec. \ref{sec:expr}. Finally, we introduce related works in Sec. \ref{sec:related_work}, and conclude in Sec. \ref{sec:conclusion}.

\section{Problem Formulation}
\label{sec:prom}

In this section, we introduce the basic settings of our data marketplace and outline our design goals.

\subsection{Preliminaries}


\textbf{Database:} Data owners contribute a database $D$ with $n$ rows, where each row $D[i]$ (i.e., a data point) corresponds to a unique data owner $u_i$. For simplicity, we assume there are $d\in Z^+$ possible values of a row, and each value is represented by a positive integer $l \in [d]$. For example, consider location databases in which $D[i]=l$ indicates that $u_i$ is at location $l$. We note that $d$ is the cardinality of a row's possible values, rather than a row's dimension.


\textbf{Linear query:} A data buyer can request a linear query $\boldsymbol{q}=(q_1,...,q_d)\in R^d$ over the database $D$. The database $D$ can be transformed into a histogram $\boldsymbol{x}=(x_1,...,x_d)$, where each element $x_l$ represents the total number of data owners whose data points are of value $l\in[d]$ (e.g., the total number of data owners who were at location $l$). Then, the raw query answer to $\boldsymbol{q}$ is the real value $y=\boldsymbol{q}(D)=\sum_{l=1}^d q_l x_l$. 

\textbf{Personalized differential privacy (PDP):} To preserve data owners' privacy, the broker cannot return raw query answers to data buyers and should thus perturb them by some perturbation mechanism $\mathcal{M}$ to achieve a formal privacy standard. Since each data owner is allowed to set a personalized upper bound of her privacy loss, we employ personalized differential privacy (PDP, a variant of differential privacy \cite{dwork2006calibrating}), where each data owner $u_i$ corresponds to a personalized privacy protection level, i.e., a privacy loss $\epsilon_i$. Jorgensen et al. \cite{jorgensen2015conservative} proposed the Sample mechanism to achieve PDP, and showed that the Laplace mechanism \cite{dwork2006calibrating} can also satisfy PDP but results in uniform privacy losses. 

\begin{Def}[Personalized Differential Privacy \cite{jorgensen2015conservative}]
\label{def:PDP}
Given privacy losses $\Phi=[\epsilon_1,...,\epsilon_n]$ where $\epsilon_i \geq 0, \forall i$, a perturbation mechanism $\mathcal{M}: \mathcal{D} \rightarrow \mathcal{R}$ satisfies $\Phi$-personalized differential privacy, if for any $u_i$, any pair of databases $D, D'$ that are neighboring with respect to $u_i$, and any possible output $o\in \mathcal{R}^d$, we have:
\begin{equation}
	Pr[\mathcal{M}(D)=o] \leq e^{\epsilon_i} Pr[\mathcal{M}(D')=o].
\end{equation}

Two databases $D, D'$ neighbor if $D'$ can be derived from $D$ by replacing one row with another, denoted as $D\sim D'$. Two databases $D, D'$ neighbor with respect to $u_i$ if $D'$ can be derived from $D$ by changing the value of $D[i]$, denoted as $D \stackrel{u_i}{\sim} D^\prime$.
\end{Def}


Intuitively, PDP guarantees that, for each data owner $u_i$, a change in the value of her data point can only cause a slight change in the perturbation mechanism's output, and such a slight change should be limited by her privacy loss $\epsilon_i$. 
Then, before we introduce PDP perturbation mechanisms, we introduce the concept of global sensitivity \cite{dwork2006calibrating}. The global sensitivity of a linear query $\boldsymbol{q}$, denoted as $\Delta\boldsymbol{q}$, is the maximum change of  $\boldsymbol{q}$ between two neighboring databases $D,D'$, i.e., $\Delta\boldsymbol{q}= \max_{D\sim D'} \left|\boldsymbol{q}(D) - \boldsymbol{q}(D')\right|=\max_l q_l - \min_{l'} q_{l'}$.

\begin{Def}[Laplace Mechanism \cite{dwork2006calibrating}]
\label{def:min_lap}
	Given a linear query $\boldsymbol{q}$, 	a database $D$, and a privacy loss $\epsilon$, the Laplace mechanism $\mathcal{L}^{\boldsymbol{q}}_{\epsilon}(D) = \mathcal{L}(D; \boldsymbol{q}, \Phi=[\epsilon,...,\epsilon])$ returns $\boldsymbol{q}(D)+\mathcal{Z}$, where $\mathcal{Z}$ is a random variable drawn from the Laplace distribution $Lap(\frac{\Delta\boldsymbol{q}}{\epsilon})$.
\end{Def}

\begin{Def}[Sample Mechanism \cite{jorgensen2015conservative}]
\label{def:sam}
Given a linear query $\boldsymbol{q}$, a database $D$, and privacy losses $\Phi=[\epsilon_1,...,\epsilon_n]$, the Sample mechanism $S^{\boldsymbol{q}}_{\Phi}$ first generates $\tilde{D}$ by sampling each row $D[i]$ with probability $\Pr_i=(e^{\epsilon_i}-1)/(e^{\max_j \epsilon_j}-1)$ and then returns $\boldsymbol{q}(\tilde{D})+ \mathcal{Z}$, where $\mathcal{Z}$ is a random variable drawn from the Laplace distribution $Lap(\frac{\Delta\boldsymbol{q}}{\max_i \epsilon_i})$.  
\end{Def}


\begin{thm}\cite{jorgensen2015conservative}
\label{thm:mechs}
The Laplace mechanism $\mathcal{L}^{\boldsymbol{q}}_{\epsilon}$ satisfies $[\epsilon,...,\epsilon]$-PDP.
The Sample mechanism $S^{\boldsymbol{q}}_{\Phi}$ satisfies $\Phi$-PDP.
\end{thm}


\textbf{Utility metric:} We use the worst-case variance as a utility metric, i.e., $v=\max_{D}Var(\tilde{y})=\max_{D}Var(\mathcal{M}(D))$. To evaluate the utility of a perturbed query answer $\tilde{y}=\mathcal{M}(D)$, i.e., a random estimation of $y$, we consider variance as a metric. We note that variance has been used as a utility metric in the proposed privacy-preserving data marketplaces \cite{li2014theory,chen2019towards}, as it reflects the expected dispersion of a random variable (i.e., a perturbed query answer in our cases). Because they employ the Laplace mechanism or its variants that add the type of noise that is independent of the raw query answer, the raw query answer cannot be revealed from the publication of the variance; however, under some perturbation mechanisms satisfying PDP, such as the Sample mechanism, data buyers can infer the raw query answer $y$ from the variance $Var(\tilde{y})$ because the value of $Var(\tilde{y})$ is dependent on the value of $D$. Therefore, the worse-case variance is more suitable for our settings.

\subsection{Market Setup}
\begin{figure}[h]
    \centering
    \includegraphics[scale=0.3]{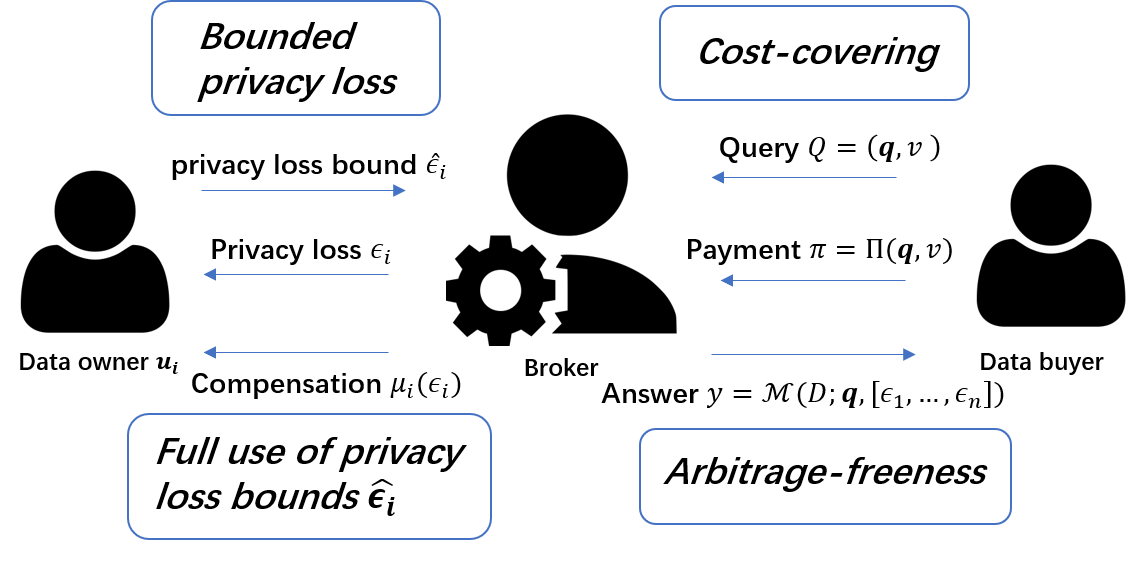}
    \caption{Market setup.}
    \label{fig:market_setup}
\end{figure}

As illustrated in Figure \ref{fig:market_setup}, there are three parties of participants in our marketplace: data owners, data buyers, and a broker (who acts as a trustworthy intermediary between data owners and data buyers). Each party has their own interests and goals. 
While data owners' privacy losses are uniform and unbounded in \cite{li2014theory}, we allow data owners to set personalized privacy loss bounds.

\textbf{Data buyers with query:} A data buyer can request one query (or more) $Q=(\boldsymbol{q},v)$ with a linear query $\boldsymbol{q}$ and a worst-case variance $v$ over the database $D$. After the buyer's payment, she obtains a perturbed query answer $\tilde{y}$ with worst-case variance $v$ perturbed by a PDP perturbation mechanism $\mathcal{M}$ with parameters $\boldsymbol{q}$ and $\Phi=[\epsilon_1,...,\epsilon_n]$, i.e., a random estimate $\tilde{y}=\mathcal{M}(D; \boldsymbol{q}, \Phi)$ such that $\max_{D} Var(\tilde{y}) =v$.


\textbf{Data owners with privacy loss:} Data owners are the source of databases stored in the marketplace. Trading a perturbed query answer $\tilde{y} = \mathcal{M}(D; \boldsymbol{q}, \Phi=[\epsilon_1,...,\epsilon_n])$ causes a privacy loss $\epsilon_i$ (which serves as a parameter of the perturbation mechanism $\mathcal{M}$) for each data owner $u_i$. Then, each $u_i$ should specify her own privacy loss bound $\hat{\epsilon}_i \in R^+$, i.e., the maximum privacy loss she can tolerate. That means, we guarantee that each data owner's privacy loss must not exceed her privacy loss bound, i.e., $\forall i, \epsilon_i \leq \hat{\epsilon}_i$.

\textbf{The broker paying compensations:} To encourage personal data trading, the broker should compensate data owners according to their privacy losses. Before data trading, each data owner $u_i$ should make a contract with the broker that the latter is obligated to compensate the former for the privacy loss $\epsilon_i$ by a compensation function $\mu_i$, which means the compensation is $\mu_i(\epsilon_i)$. Thus, the cost of trading a perturbed query answer $\tilde{y}$ is the sum of the compensations to data owners, defined as $\boldsymbol{\mu}([\epsilon_1,...,\epsilon_n])=\sum_i \mu_i(\epsilon_i)$, where $\boldsymbol{\mu}$ is the cost function that takes a vector of privacy losses as input and outputs the cost. We reasonably assume that each compensation function $\mu_i$ is an increasing and twice differentiable function and $\mu_i(0)=0$.

\textbf{The broker setting query prices:} As a profit-making intermediary, the broker sells perturbed query answers to make a profit. Because data buyers care more about the utility of query answers, we set that the price of a perturbed query answer directly depends on its worst-case variance $v$, rather than on the privacy losses it caused. The broker should design a \textit{pricing function} $\pi=\Pi(\boldsymbol{q},v)$ that is (1) \textit{cost recovering} (see Definition \ref{def:cost recovering}), which means that a query price should cover the total compensations to data owners, and (2) \textit{arbitrage free} (see Definition \ref{def:ab}), which means that a data buyer cannot obtain a perturbed query answer with a specific worst-case variance $v$ more cheaply by deriving it from a less expensive set of perturbed query answers. 

\begin{Def}[Cost Recovering]
\label{def:cost recovering}
A pricing function $\pi=\Pi(\boldsymbol{q},v)$ ($v>0$) is considered to be cost recovering if it satisfies the following: for any perturbed query answer $\tilde{y}$ with a worst-case variance $v$ that caused a privacy loss $\epsilon_i$ for each data owner $u_i$, we have $\pi=\Pi(\boldsymbol{q},v) \geq \boldsymbol{\mu}([\epsilon_1,...,\epsilon_n]) = \sum_i \mu_i(\epsilon_i)$. 
\end{Def}

\begin{Def}[Arbitrage Freeness]
\label{def:ab}
A pricing function $\pi=\Pi(\boldsymbol{q}, v)$ is considered to be arbitrage free if it satisfies the following: for every multiset of pairs of a linear query and a worst-case variance $\{(\boldsymbol{q}_1,v_1),...,(\boldsymbol{q}_m,v_m)\}$ where $m \in N^+$, if there exists $a_1,...,a_m$ such that $\sum_{j=1}^m a_j \boldsymbol{q}_j=\boldsymbol{q}$ and $\sum_{j=1}^m a_j^2v_j\leq v$, we have $\Pi(\boldsymbol{q},v)\leq \sum_{j=1}^m\Pi(\boldsymbol{q}_j,v_j)$. 
\end{Def}

\textbf{Design goals:} We clarify the design goals of our data market that empowers data owners with the control of privacy loss and avoids arbitrage: 
\begin{itemize}[leftmargin=*]
	\item \textit{Data owners:} We guarantee that each data owner's privacy loss is controlled within her privacy loss bound. In the meantime, the design goal for data owners is to make full use of their privacy loss bounds, which means we consume their tolerable privacy losses as much as possible so that they can gain higher compensation.
	\item \textit{The broker:} We guarantee that the pricing function is cost recovering (see Definition \ref{def:cost recovering}) and arbitrage free (see Definition \ref{def:ab}). However, in Sec. \ref{sec:partial_ab}, we relax the guarantee of arbitrage freeness and instead guarantee partial arbitrage freeness (see Definition \ref{def:partial_ab}).
\end{itemize} 

\section{Personal Data Trading Framework}
\label{sec:framework}

In this section, we first give a short view of our personal data trading framework with two key modules. Then, we discuss more details about the design of arbitrage-free pricing in our framework. Finally, we present how our trading framework supports partial arbitrage freeness. In summary, we provide a framework to compute arbitrage-free trading protocols. Some important notations are summarized in Table \ref{table:notations}.

\begin{table}[t]
\caption{Summary of Notations}
\centering
\begin{spacing}{1}
	\begin{tabular}{|c|}
	\hline
		Notation: Description\\	
	\hline
		$u_i$: data owner $i$\\
	\hline
		$n$: total number of users\\
	\hline
		$D$: database\\
	\hline
		$Q=(\boldsymbol{q},v)$: a query $Q$ with a linear query $\boldsymbol{q}$ over $D$\\
		and a worst-case variance $v$\\
	\hline
		$\tilde{y}$: perturbed query answer to $Q$\\
	\hline
		$\check{v}$: lower bound of $v$\\
	\hline
		$\hat{\epsilon}_i$: $u_i$'s privacy loss bound\\
	\hline
		$\bar{\epsilon}_i$: $u_i$'s privacy budget allocated for $Q$\\	
	\hline
		$\epsilon_i$: $u_i$'s privacy loss for $Q$\\
	\hline
		$\mu_i(\cdot)$: $u_i$'s compensation function\\
        \hline
		$\boldsymbol{\mu}(\cdot)$: cost function, $\boldsymbol{\mu}([\epsilon_1,...,\epsilon_n])=\sum_i \mu_i(\epsilon_i)$\\
	\hline
		$\mathcal{M}$: perturbation mechanism\\
	\hline	
		$v=U_{\mathcal{M}}(\boldsymbol{q},[\epsilon_1,...,\epsilon_n])$: utility function under $\mathcal{M}$\\ 
	\hline
		$\pi=\Pi_{\mathcal{M}}(\boldsymbol{q},v)$: pricing function under $\mathcal{M}$\\
	\hline
\end{tabular}
\end{spacing}
\label{table:notations}
\end{table}

\begin{figure}[ht]
\centerline{\includegraphics[scale=0.25]{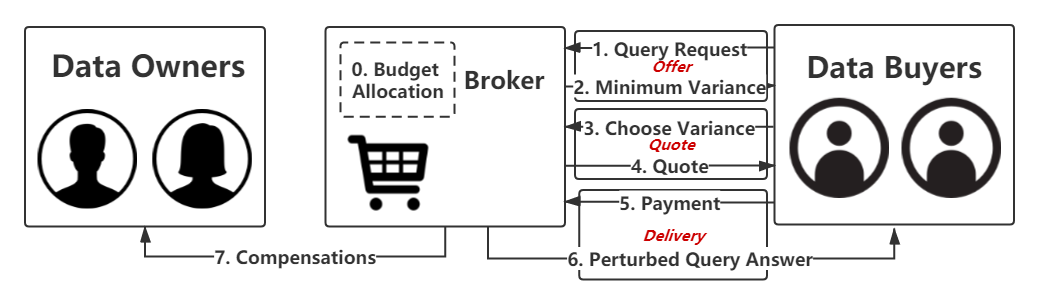}}
\caption{Transaction Flow in our Trading Framework.}
\label{fig:trading_framework}
\end{figure}
\subsection{Framework Overview}

 As depicted in Figure \ref{fig:trading_framework} and Alg. \ref{alg:framework}, a transaction of a query in our data marketplace works in the following manner.


\textbf{Budget allocation:}
In Step $0$, the broker should allocate a part of data owners' privacy loss bounds $[\hat{\epsilon}_1,...,\hat{\epsilon}_n]$ for the coming query as privacy budgets $[\bar{\epsilon}_1,...,\bar{\epsilon}_n]$ (i.e., the budgets of their privacy losses $[\epsilon_1,...,\epsilon_n]$). This means that their privacy losses to be leaked from the coming query should be no higher than the privacy budgets, i.e., $\forall i, \epsilon_i \leq \bar{\epsilon}_i \leq \hat{\epsilon}_i$.

\textbf{Offer:} In Step $1$, a data buyer requests a linear query $\boldsymbol{q}$ over database $D$. Then, in Step $2$, the broker returns the maximum available utility to the data buyer, i.e., the lower bound $\check{v}$ of worst-case variance $v$. Here, we define the utility function of query answers perturbed by $\mathcal{M}$ as $v=U_{\mathcal{M}}(\boldsymbol{q},[\epsilon_1,...,\epsilon_n])$, which takes as input the pair of a linear query $\boldsymbol{q}$ and a vector of privacy losses $[\epsilon_1,...,\epsilon_n]$ and outputs the worst-case variance $v=\max_D Var(\mathcal{M}(D;\boldsymbol{q},[\epsilon_1,...,\epsilon_n]))$. The value of $\check{v}$ depends on the linear query $\boldsymbol{q}$ and privacy budgets $[\bar{\epsilon}_1,...,\bar{\epsilon}_n]$, i.e., $\check{v}=U_{\mathcal{M}}(\boldsymbol{q},[\bar{\epsilon}_1,...,\bar{\epsilon}_n])$.


\textbf{Quote:} In Step $3$, given the lower bound $\check{v}$, the data buyer should choose a worst-case variance $v \geq \check{v}$. Then, in Step $4$, the broker quotes a price $\pi=\Pi_\mathcal{M}(\boldsymbol{q},v)$ for the perturbed query answer to $Q=(\boldsymbol{q},v)$ using an arbitrage-free and cost-recovering pricing function $\Pi_\mathcal{M}$. 

\textbf{Delivery:} In Step $5$, the data buyer pays the price $\pi$. Then, in Step $6$, the broker calculates data owners' privacy losses $[\epsilon_1,...,\epsilon_n]$ and returns the perturbed query answer $\tilde{y}=\mathcal{M}(D; \boldsymbol{q}, [\epsilon_1,...,\epsilon_n])$ to the buyer. Finally, in Step $7$, the broker compensates each data owner $u_i$ according to her compensation function $\mu_i$ and privacy loss $\epsilon_i$. 

\subsection{Key Modules}



Here, we introduce two key modules in our framework (Alg. \ref{alg:framework}), \textit{PerturbMech} and \textit{BudgetAlloc}. As shown in Alg. \ref{alg:framework}, before trading query answers over a database, the broker should make a contract with each data owner $u_i$ to stipulate the compensation function $\mu_i$ and the privacy loss bound $\hat{\epsilon}_i$. In addition, the broker should decide which instances of the following key modules to deploy. Briefly, the \textit{BudgetAlloc} module is designed for privacy budget allocation, and \textit{PerturbMech} is designed for perturbing query answers to satisfy personalized differential privacy.

\begin{algorithm}[h]
\caption{Arbitrage-Free Trading Framework}
\label{alg:framework}
\begin{algorithmic}[1]
\REQUIRE privacy loss bounds $\hat{\Phi}=[\hat{\epsilon}_1,...,\hat{\epsilon}_n]$, subadditive compensation functions $[\mu_1,...,\mu_n]$

\WHILE {$\forall i, \hat{\epsilon}_i > 0$}
	\STATE $[\bar{\epsilon}_1,...,\bar{\epsilon}_n] \gets \textbf{BudgetAlloc}([\hat{\epsilon}_1,...,\hat{\epsilon}_n])$ // Step 0
	\STATE Receive a linear query $\boldsymbol{q}$ // Step 1
	\STATE $\check{v} \gets U_{\mathcal{M}}(\boldsymbol{q}, [\bar{\epsilon}_1,...,\bar{\epsilon}_n])$ // Step 2
	\STATE Receive a worst-case variance $v$ // Step 3
	\IF{$v<\check{v}$}
		\STATE Reject $Q=(\boldsymbol{q}, v)$
	\ELSE
		\STATE Quote the buyer $\pi=\Pi_{\mathcal{M}}(\boldsymbol{q}, v)$ // Step 4
		\STATE Collect the buyer's payment // Step 5
		\STATE $[\epsilon_1,...,\epsilon_n] \gets {U^{\boldsymbol{q}}_{\mathcal{M}}}^{-1}(v)$
		\STATE $\tilde{y} \gets \textbf{PerturbMech}(D; \boldsymbol{q}, [\epsilon_1,...,\epsilon_n])$
		\STATE Deliver $\tilde{y}^\tau$ to the buyer // Step 6
		\STATE $\hat{\epsilon}_i \gets \hat{\epsilon}_i - \bar{\epsilon}_i, \forall i$
		\STATE Compensate each $u_i$ by $\mu_i(\epsilon_i)$ // Step 7
	\ENDIF
\ENDWHILE
\end{algorithmic}
\end{algorithm}

\begin{figure}[h]
    \centering
    \includegraphics[scale=0.5]{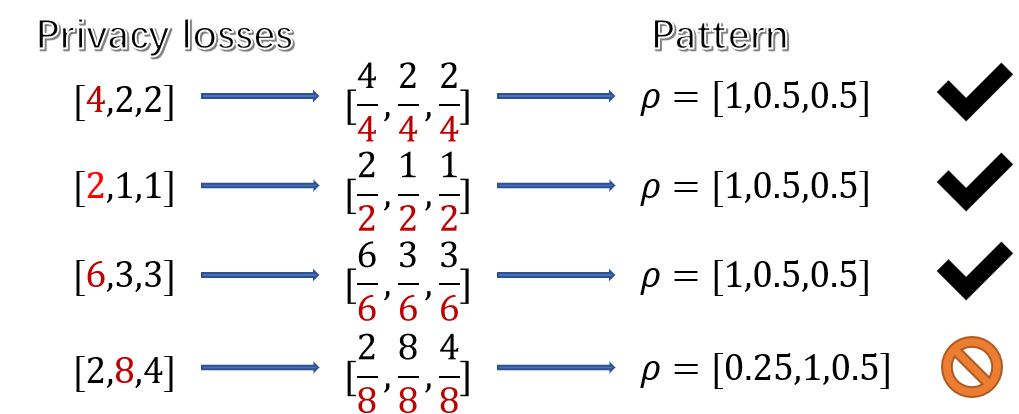}
    \caption{$\rho$-pattern.}
    \label{fig:pattern}
\end{figure}

\textbf{BudgetAlloc():} This module takes as input data owners' (remaining) privacy loss bounds $[\hat{\epsilon}_1,...,\hat{\epsilon}_n]$ and outputs their privacy budgets $[\bar{\epsilon}_1,...,\bar{\epsilon}_n]$ ($\epsilon_i \leq \bar{\epsilon}_i \leq \hat{\epsilon}_i, \forall i$) for the coming query in a way that guarantees arbitrage freeness. 
Although data owners authorize the broker to utilize their privacy losses within their privacy loss bounds to perturb query answers, due to the constraint of arbitrage freeness, the broker cannot arbitrarily spend those privacy loss bounds.
Instead, to satisfy arbitrage freeness, in our framework, the broker always consumes their privacy losses $[\epsilon_1,...,\epsilon_n]$ in a given $\rho$-pattern (see Definition \ref{def:pattern}), which means the ratios between every two data owners' privacy losses $\epsilon_i, \epsilon_j, \forall i, j \in [n]$ are fixed. 
For example, in Figure \ref{fig:pattern}, only the privacy losses in the given $[1, 0.5, 0.5]$-pattern are allowed. 
Hence, to spend privacy losses in a given $\rho$-pattern, \textit{BudgetAlloc} always outputs privacy budgets in the $\rho$-pattern, i.e., $[\bar{\epsilon}_1,...,\bar{\epsilon}_n]=\rho \cdot \max_i \bar{\epsilon}_i$. 
In Sec. \ref{sec:ab_pricing}, we will explain why and how we use an appropriate pattern $\rho$ to guarantee arbitrage freeness.

\begin{Def}[$\rho$-Pattern]
\label{def:pattern}
A pattern is a vector $\rho=[\rho_1,...,\rho_n]$ where each pattern element $\rho_i \in \rho$ satisfies $0 \leq \rho_i \leq 1$ and $\exists j \in [n]$ such that $\rho_j=1$. If a vector $[\epsilon_1,...,\epsilon_n]$ is equal to $\rho \cdot \max_i \epsilon_i$, then the vector is in the $\rho$-pattern. We also use $\theta$ to denote $\max_i \epsilon_i$. 
\end{Def}

\textbf{PerturbMech():} This module aims at perturbing query answers to satisfy personalized differential privacy (PDP). 
An instance of this module should be a PDP perturbation mechanism $\mathcal{M}$, such as the Laplace mechanism and the Sample mechanism. 
In addition, in Sec. \ref{sec:ab_pricing}, we will present that the utility function should satisfy the second and third properties in Theorem \ref{theorem:arbitrage free} to guarantee arbitrage freeness. 
Since whether these two properties can be satisfied naturally depends on the selection of the perturbation mechanism $\mathcal{M}$, we should carefully design or select a perturbation mechanism $\mathcal{M}$ achieving not only PDP but also these properties.

\subsection{Method for Ensuring Arbitrage-free Pricing}
\label{sec:ab_pricing}

Here, we discuss in detail the design of arbitrage-free pricing in our framework. 

\textbf{Existing techniques:}
Li et al. \cite{li2014theory} proposed a method for designing arbitrage-free pricing functions in cases where the privacy losses are uniform and the perturbation mechanism is not complex. 
Concretely, when data owners' privacy losses are uniform as $\epsilon$, given the utility function $v=U_{\mathcal{M}}(\boldsymbol{q}, [\epsilon,...,\epsilon])$ of the perturbation mechanism $\mathcal{M}$, for each data owner $i$, they first derive the inverse utility function $\epsilon = {U^{\boldsymbol{q}}_{\mathcal{M}}}^{-1}(v)$ and construct a micro-payment function $\Pi_{\mathcal{M},i}(\boldsymbol{q}, v) = \mu_i( {U^{\boldsymbol{q}}_{\mathcal{M}}}^{-1}(v))$; 
if $\Pi_{\mathcal{M},i}(\boldsymbol{q}, v)$ can be identified as one of certain composite functions of arbitrage-free pricing functions, then it is also arbitrage free; 
finally, if $\Pi_{\mathcal{M},i}(\boldsymbol{q}, v)$ is arbitrage free for all $i$,
they can derive an arbitrage-free pricing function $\Pi_{\mathcal{M}}(v) = \sum_i \Pi_{\mathcal{M},i}(\boldsymbol{q}, v)$.

\textbf{Challenges:}
However, the method proposed by Li et al. \cite{li2014theory} cannot work in our framework due to the employment of personalized differential privacy (PDP). First, because the privacy losses $\epsilon_1,...,\epsilon_n$ are personalized, the inverse utility functions $\epsilon_i= {U^{\boldsymbol{q}}_{\mathcal{M}, i}}^{-1}(v), \forall i$ (i.e., the mapping from the worst-case variance $v$ to data owner $i$' privacy loss) may not exist for constructing the micro-payment functions $\Pi_{\mathcal{M},i}(\boldsymbol{q}, v) = \mu_i ({U^{\boldsymbol{q}}_{\mathcal{M}, i}}^{-1}(v)), \forall i$. Second, even if the inverse utility functions exist, under a sophisticated PDP perturbation mechanism (e.g., the Sample mechanism \cite{jorgensen2015conservative}), the utility function can be so complex that the inverse utility functions cannot be expressed in analytic form, which makes it extremely difficult to identify whether the micro-payment functions are arbitrage free or not. 

\textbf{Our approach:}
Hence, we propose a general method for guaranteeing arbitrage freeness. First, we also utilize the inverse utility function to construct a pricing function. Then, we use the proposed notion $\rho$-Pattern to ensure the existence of the inverse utility function. Finally, we guarantee arbitrage freeness by satisfying certain properties of the utility function (rather than the inverse utility function).



\textbf{1. Constructing a pricing function:}
\textit{First, to ensure that the price can cover all the compensations to data owners, we construct a pricing function by the inverse utility function.} Let $v=U^{\boldsymbol{q}}_\mathcal{M}([\epsilon_1,...,\epsilon_n])$ be the univariate utility function satisfying $ U^{\boldsymbol{q}}_\mathcal{M}([\epsilon_1,...,\epsilon_n]) \equiv U_\mathcal{M}(\boldsymbol{q}, [\epsilon_1,...,\epsilon_n])$. Given a query $(\boldsymbol{q}, v)$, to make its price cost recovering, we can sum up the corresponding compensations to data owners as the pricing function. That is, we can derive the corresponding privacy losses by the inverse utility function $[\epsilon_1,...,\epsilon_n]={U^{\boldsymbol{q}}_\mathcal{M}}^{-1}(v)$ and then construct a cost-recovering pricing function $\pi=\Pi_\mathcal{M}(\boldsymbol{q},v)=\boldsymbol{\mu}([\epsilon_1,...,\epsilon_n]) = \boldsymbol{\mu}({U^{\boldsymbol{q}}_\mathcal{M}}^{-1}(v))$. We note that a profit-making broker can simply modify this pricing function to derive another one that covers a profit, e.g., $\pi=\Pi_\mathcal{M}(\boldsymbol{q},v)=(1+r)\boldsymbol{\mu}({U^{\boldsymbol{q}}_\mathcal{M}}^{-1}(v))$ where $r>0$ is a flat profit rate; we employ $\pi=\Pi_\mathcal{M}(\boldsymbol{q},v)=\boldsymbol{\mu}({U^{\boldsymbol{q}}_\mathcal{M}}^{-1}(v))$ for simplicity, but our techniques support the former case.

\textbf{2. Ensuring the inverse utility function's existence:}
\textit{Then, to ensure the existence of the inverse utility function, we utilize a pattern for budget allocation.} 
Under a simple perturbation mechanism with uniform privacy losses, such as the Laplace mechanism \cite{dwork2006calibrating}, the inverse utility function ${U^{\boldsymbol{q}}_\mathcal{M}}^{-1}$ can be easy to derive. However, as we mentioned above, ${U^{\boldsymbol{q}}_\mathcal{M}}^{-1}$ may not exist under a complex perturbation mechanism with personalized privacy losses, such as the Sample mechanism \cite{jorgensen2015conservative}. To guarantee the existence of ${U^{\boldsymbol{q}}_\mathcal{M}}^{-1}$, first, we introduce the notion $\rho$-Pattern to constrain the domain of the utility function's input. Intuitively, the inverse function does not exist because an output of the utility function corresponds to multiple inputs. Hence, we can rule out redundant inputs by fixing the pattern of valid inputs so that there is a bijection between the sets of worst-case variance and privacy losses. 
As depicted in Figure \ref{fig:pattern}, we can use a fixed pattern $\rho$ to constrain the domain of the utility function so that every input $[\epsilon_1,...,\epsilon_n]$ is in the $\rho$-pattern, i.e., $[\epsilon_1,...,\epsilon_n]=\rho \cdot \theta$. That also means that for each query $Q$, the privacy budgets or privacy losses should be allocated or consumed in a fixed $\rho$-pattern, which makes $U_\mathcal{M}(\boldsymbol{q}, [\epsilon_1,...,\epsilon_n])=U_\mathcal{M}(\boldsymbol{q}, \rho\cdot \theta)$ hold.
Thus, the pricing function is derived as follows:
\begin{equation*}
\pi=\Pi_{\mathcal{M}}(\boldsymbol{q},v)= \boldsymbol{\mu}({U^{\boldsymbol{q}}_\mathcal{M}}^{-1}(v))=\boldsymbol{\mu}(\rho \cdot \theta).
\end{equation*}
Then, we guarantee the existence of ${U^{\boldsymbol{q}}_\mathcal{M}}^{-1}$ by finding a $\rho$ such that ${U^{\boldsymbol{q}}_{\mathcal{M}}}(\rho\cdot\theta)$ decreases as $\theta$ increases, which is also implied by the fourth property in Theorem \ref{theorem:arbitrage free}.


\begin{Rem}
You may argue that selling privacy losses in a fixed pattern $\rho$ is unfair because a data owner $u_i$ with a higher pattern element $\rho_i$ can sell a higher privacy loss for each query. In Remark \ref{rem:pattern_search}, we show that this unfair design can be perceived as an incentive that encourages data owners to set a higher privacy loss bound.
\end{Rem}

\textbf{3. Ensuring arbitrage freeness:}
\textit{Finally, to ensure arbitrage freeness when the inverse utility function is hard to analyze, we guarantee certain properties of the utility function.}
As mentioned before, under some perturbation mechanisms $\mathcal{M}$, the utility function can be so complex that the inverse function ${U^{\boldsymbol{q}}_\mathcal{M}}^{-1}$ cannot even be expressed in analytic form, which makes it difficult to analyze the mathematical properties of the pricing function $\Pi_{\mathcal{M}}$ and challenging to guarantee its arbitrage freeness. To solve this challenge, we consider whether we can instead study the properties of the utility function $U_{\mathcal{M}}$ to guarantee arbitrage-free pricing and propose Theorem \ref{theorem:arbitrage free}, illustrating that if $U_{\mathcal{M}}$ satisfies certain properties, $\Pi_{\mathcal{M}}$ is arbitrage free.


\begin{thm}
\label{theorem:arbitrage free}
Let each data owner $u_i$'s compensation function $\mu_i$ be subadditive. The pricing function $\pi=\Pi_{\mathcal{M}}(\mathbf{q}, v)=\boldsymbol{\mu}({U^{\mathbf{q}}_{\mathcal{M}}}^{-1}(v))$ is arbitrage free if the utility function $v=U_{\mathcal{M}}(\mathbf{q}, [\epsilon_1,...,\epsilon_n])$ satisfies the following properties:
\begin{enumerate}
\item there exists a variable $\theta > 0$, such that $\forall i$, there is an increasing, subadditive and twice differentiable function $f_i:R\rightarrow R$ such that $\epsilon_i=f_i(\theta)$ and $f_i(0)=0$,
\item  there exists a function $\eta: R^n \rightarrow R$ and a seminorm $\lambda: R^d \rightarrow R$ such that $U_{\mathcal{M}}(\mathbf{q}, [\epsilon_1,...,\epsilon_n])\equiv\lambda^2(\mathbf{q})\cdot \eta([\epsilon_1,...,\epsilon_n])$,
\item $\forall \boldsymbol{q}\neq \boldsymbol{0},\lim_{\theta \rightarrow 0^+} U_{\mathcal{M}}(\mathbf{q}, [\epsilon_1,...,\epsilon_n]) = +\infty$,
\item $\forall \theta > 0$, $\frac{\partial v}{\partial \theta}<0$, and
\item $\forall \theta > 0$, $v \cdot \frac{\partial^2 v}{\partial \theta^2}\leq 2(\frac{\partial v}{\partial \theta})^2$.
\end{enumerate}
\end{thm}

\begin{proofnoof}
All the missing proofs are given in the appendix.
\end{proofnoof}
Thus, we can transform the problem of designing an arbitrage-free pricing function into finding a utility function satisfying such properties. 
Concretely, \textit{PerturbMech} should guarantee the second property and \textit{BudgetAlloc} the others.
In fact, the second property in Theorem \ref{theorem:arbitrage free} is satisfied under the Laplace mechanism or the Sample mechanism because the global sensitivity of the linear query is a seminorm $\lambda$, i.e., $\lambda(\boldsymbol{q}) = \Delta\boldsymbol{q}=\max_{l, l'} |q_l - q_{l'}|$. 
Then, if we make every vector of privacy losses $[\epsilon_1,...,\epsilon_n]$ in a $\rho$-pattern, the first property is also satisfied by the utility function $v=U_{\mathcal{M}}(\mathbf{q}, \rho\cdot \theta)$ because for each $i$, there exists an increasing, subadditive and twice differentiable function $\epsilon_i = f_i (\theta) = \rho_i \cdot \theta$.
Finally, the selection of $\rho$ is essentially the key to arbitrage freeness because a well-tuned $\rho$ can ensure the remaining properties. 

\begin{Rem}[Intuition behind Theorem \ref{theorem:arbitrage free}]
\label{rem:arbitrage free}
Here, we explain the properties in Theorem \ref{theorem:arbitrage free}. The first property requires that the relationships between elements of $[\epsilon_1,...,\epsilon_n]$ should remain unchanged throughout all rounds of data trading; if $[\epsilon_1,...,\epsilon_n]$ is always allocated in a $\rho$-pattern, there exists a linear relationship between every two privacy losses $\epsilon_i,\epsilon_j \in [\epsilon_1,...,\epsilon_n]$. The second property is proposed to prevent arbitrage behaviors combing multiple queries $(\boldsymbol{q}, v)$ with different $\boldsymbol{q}$ but $v$ held constant, while the rest properties guarantee arbitrage freeness in the aspect of combing queries with different $v$ but $\boldsymbol{q}$ held constant. Concretely, the third means zero privacy losses correspond to an infinitely high worst-case variance, which implies that no useful query answer can be traded for free. The fourth makes the worst-case variance $v$ decrease as $\theta$ increases, so that $\Pi_{\mathcal{M}}$ is a monotonically decreasing function with respect to $v$; the fifth ensures a slow decreasing speed of $\Pi_{\mathcal{M}}$ with respect to $v$, so that purchasing a query answer with a lower $v$ is more worthwhile than with a higher $v$. In addition, if the second property is satisfied, the inverse function $\eta^{-1}$ is equivalent to the inverse function ${U^{\boldsymbol{q}}_\mathcal{M}}^{-1}$, i.e., $[\epsilon_1,...,\epsilon_n]={U^{\boldsymbol{q}}_\mathcal{M}}^{-1}(v)=\eta^{-1}(\frac{v}{\lambda^2(\boldsymbol{q})})$. Then, the fourth property guarantees the existence of $\eta^{-1}$ and ${U^{\boldsymbol{q}}_\mathcal{M}}^{-1}$.
\end{Rem}

\subsection{Partial Arbitrage Freeness}
\label{sec:partial_ab}
\textbf{Motivation:} Partial arbitrage freeness (see Definition \ref{def:partial_ab}) is a reasonable relaxation of arbitrage freeness that naturally fits our settings of bounded utilities. In the previous subsection, we have shown the complexity of guaranteeing arbitrage freeness. However, because the privacy loss is bounded, the utility of the perturbed query answer is also bounded. Therefore, it is reasonable to only guarantee arbitrage freeness partially for a limited range of utility, i.e., to guarantee partial arbitrage freeness.

\begin{Def}[Partial Arbitrage Freeness]
\label{def:partial_ab}
A pricing function $\pi=\Pi(\boldsymbol{q}, v)$ is partially arbitrage free for $v\in[\psi^{L}(\boldsymbol{q}),\psi^{U}(\boldsymbol{q})]$ where $\psi^{U}(\boldsymbol{q}) \geq \psi^{L}(\boldsymbol{q}) \geq 0$, if it satisfies the following condition: for every possible multiset of pairs of a linear query and a worst-case variance $\{(\boldsymbol{q}_1,v_1),...,(\boldsymbol{q}_m,v_m)\}$ where $m \in Z^+$ and $\forall j\in [m], v_j\in [\psi^{L}(\boldsymbol{q}),\psi^{U}(\boldsymbol{q})]$, if there exists $a_1,...,a_m$ such that $\sum_{j=1}^m a_j \boldsymbol{q}_j=\boldsymbol{q}$ and $\sum_{j=1}^m a_j^2v_j\leq v$, we have $\Pi(\boldsymbol{q},v)\leq \sum_{j=1}^m\Pi(\boldsymbol{q}_j,v_j)$. 
\end{Def}

\begin{Rem}
The partially arbitrage-free range $[\psi^{L}(\boldsymbol{q}),\psi^{U}(\boldsymbol{q})]$ of the worst-case variance $v$ has a factor of the linear query $\boldsymbol{q}$ because $\boldsymbol{q}$ also affects the properties of the pricing function $\pi=\Pi(\boldsymbol{q}, v)$. If $\boldsymbol{q}$ is given, then the partially arbitrage-free range is also determined.
\end{Rem}

\textbf{Cases calling for partial arbitrage freeness:}  In some cases, it is not necessary to guarantee arbitrage freeness for the whole range of queries $Q=(\boldsymbol{q},v)$, e.g., for those that cannot be answered or seldom requested by data buyers.
\begin{itemize}[leftmargin=*]
\item \textbf{Queries with low utility:} If those queries with extremely high worst-case variance are seldom requested because they are relatively too inaccurate, we can only guarantee partial arbitrage freeness for $v\leq  U_{\mathcal{M}}(\boldsymbol{q}, \rho \cdot \theta^{L})$, where $\theta^{L} > 0$ is a constant set by the broker.
\item \textbf{Unanswerable queries:} Given a linear query $\boldsymbol{q}$, the worst-case variance $v$ requested by the data buyer should be no lower than $\check{v}$, we only need to ensure that the pricing function $\pi = \Pi_{\mathcal{M}}(\boldsymbol{q},v)$ is partially arbitrage free for $v\geq \check{v} = U_{\mathcal{M}}(\boldsymbol{q}, [\bar{\epsilon}_1,...,\bar{\epsilon}_n])$. 

\end{itemize} 
Hence, we propose the following Theorem \ref{thm:partial_ab}, an extension of Theorem \ref{theorem:arbitrage free}, to achieve partial arbitrage freeness, which shows that if the utility function $U_{\mathcal{M}}$ satisfies some properties, the pricing function $\Pi_\mathcal{M}$ is partially arbitrage free for some range of queries.

\begin{thm}
\label{thm:partial_ab}
Let each data owner $u_i$'s compensation function $\mu_i$ be subadditive. Given two constants $\theta^{U} \geq\theta^L > 0$, the pricing function $\pi=\Pi_{\mathcal{M}}(\mathbf{q}, v)=\boldsymbol{\mu}({U^{\mathbf{q}}_{\mathcal{M}}}^{-1}(v))$ is partially arbitrage free for $v\in [U_{\mathcal{M}}(\boldsymbol{q},[f_1(\theta^{U}),...,f_n(\theta^{U})]),$ $U_{\mathcal{M}}(\boldsymbol{q},[f_1(\theta^{L}),...,f_n(\theta^{L})])]$ if the utility function $v=U_{\mathcal{M}}(\mathbf{q}, [\epsilon_1,...,\epsilon_n])$ satisfies the following properties:
\begin{enumerate}
\item there exists a variable $\theta > 0$, such that $\forall i$, there is an increasing, subadditive and twice-differentiable function $f_i:R\rightarrow R$ such that $\epsilon_i=f_i(\theta)$ and $f_i(0)=0$,
\item  there exists a function $\eta: R^n \rightarrow R$ and a seminorm $\lambda: R^d \rightarrow R$ such that $U_{\mathcal{M}}(\mathbf{q}, [\epsilon_1,...,\epsilon_n])\equiv\lambda^2(\mathbf{q})\cdot \eta([\epsilon_1,...,\epsilon_n])$,
\item $\forall \theta \geq \theta^L, \theta \leq \theta^U -\theta^L, U_{\mathcal{M}}(\boldsymbol{q},[f_1(\theta + \theta^{L}),...,f_n(\theta+ \theta^{L})])\leq
\frac{1}{\frac{1}{U_{\mathcal{M}}(\boldsymbol{q},[f_1(\theta^{L}),...,f_n(\theta^{L})])} + \frac{1}{U_{\mathcal{M}}(\boldsymbol{q},[f_1(\theta),...,f_n(\theta)])}}$,
\item  $\forall \theta > 0$, $\frac{\partial v}{\partial \theta}<0$, and
\item  $\forall \theta \in [\theta^{L}, \theta^{U}]$, $v\cdot \frac{\partial^2 v}{\partial \theta^2}\leq 2(\frac{\partial v}{\partial \theta})^2$.
\end{enumerate}
\end{thm}

\begin{Rem}
Here, we explain the five properties in Theorem \ref{thm:partial_ab}. We denote $[f_1(\theta),...,f_n(\theta)]$ as $[f_i(\theta)]$. 
\begin{itemize}[leftmargin=*]
\item The first, second and fourth properties in Theorem \ref{thm:partial_ab} and Theorem \ref{theorem:arbitrage free} are the same. 
\item The third property requires an appropriate value of the endpoint $U_{\mathcal{M}}(\boldsymbol{q},[f_i(\theta^{L})])$. If a data buyer purchases two queries $(\boldsymbol{q}, v_1)$ and $(\boldsymbol{q}, v_2)$ to commit arbitrage, the optimal strategy is to combine their answers $\tilde{y}_1, \tilde{y}_2$ into a query answer $\tilde{y}_3 = \frac{v_2}{v_1 + v_2}\tilde{y}_1 + \frac{v_1}{v_1 + v_2}\tilde{y}_2$ so that the worst-case variance is minimized, i.e., $v_3 = 1/(\frac{1}{v_1} + \frac{1}{v_2})$. 
Let $v_1 = U_{\mathcal{M}}(\boldsymbol{q}, [f_i(\theta^{L})])$ and $v_2 = U_{\mathcal{M}}(\boldsymbol{q}, [f_i(\theta)])$.
Then, the minimized worst-case variance $v_3$ is the right side of the inequality, which is no lower than $v_4=U_{\mathcal{M}}(\boldsymbol{q},[f_i(\theta+\theta^{L})])$. Then, because $\mu_i$ and $f_i$ are subadditive for any $i$, the total price $\boldsymbol{\mu}([f_i(\theta^{L})]) + \boldsymbol{\mu}([f_i(\theta)])$ of purchasing both $\tilde{y}_1$ and $\tilde{y}_2$ must be no lower than the price $\boldsymbol{\mu}([f_i(\theta+\theta^{L})])$ of query $(\boldsymbol{q}, v_4)$. Therefore, arbitrage against $(\boldsymbol{q},v_4)$ cannot occur by purchasing $\tilde{y}_1, \tilde{y}_2$.

\item The fourth and fifth properties require that the utility function decreases rapidly as $\theta$ increases when $\theta$ is in the range of $ [\theta^{L}, \theta^{U}]$, so that the pricing function decreases slowly as $v$ increases when $v$ is in the partially arbitrage-free range $[U_{\mathcal{M}}(\boldsymbol{q},[f_i(\theta^{U})]),  U_{\mathcal{M}}(\boldsymbol{q},[f_i(\theta^{L})])]$. 

\item The last three properties guarantee partial arbitrage freeness with $\boldsymbol{q}$ held constant. Concretely, arbitrage cannot occur by purchasing any possible $m\in N^+$ queries within the partially arbitrage-free range because those properties imply that 
\begin{equation*}
\begin{aligned}
&\forall \theta_1, ..., \theta_m \geq \theta^L, \sum_{j=1}^{m}\theta_j \leq \theta^U,\\
&U_{\mathcal{M}}(\boldsymbol{q},[f_i(\sum_{j=1}^{m} \theta_j)]) \leq 1/\sum_{j=1}^{m} \frac{1}{U_{\mathcal{M}}(\boldsymbol{q},[f_i(\theta_j)])},
\end{aligned}
\end{equation*}
which is a generalization of the third property.
\end{itemize}
\end{Rem}

 \begin{algorithm}[ht]
\caption{Partially Arbitrage-Free Trading Framework}
\label{alg:pab_framework}
\begin{algorithmic}[1]
\REQUIRE privacy loss bounds $\hat{\Phi}=[\hat{\epsilon}_1,...,\hat{\epsilon}_n]$, compensation functions $[\mu_1,...,\mu_n]$, $\theta^L$, $\theta^{U}$
\WHILE{$\forall i, \hat{\epsilon}_i > 0$}
\STATE Same as Line 2 to Line 5 of Algorithm \ref{alg:framework}
	\IF{$v \notin [\check{v}, +\infty) \cap  [U_{\mathcal{M}}(\boldsymbol{q},\rho\cdot\theta^{U}),  U_{\mathcal{M}}(\boldsymbol{q},\rho\cdot\theta^{L})]$}
			\STATE Reject $Q=(\boldsymbol{q}, v)$
      \ELSE
      		\STATE Same as Line 9 to Line 15 of Algorithm \ref{alg:framework}
         \ENDIF
\ENDWHILE
\end{algorithmic}
\end{algorithm}

\textbf{Modified framework (Alg. \ref{alg:pab_framework}):} Our partially arbitrage-free trading framework generally stays the same as our arbitrage-free trading framework (Alg. \ref{alg:framework}), except the following slight modifications.

\begin{itemize}[leftmargin=*]
\item The broker only answers those answerable queries $Q=(\boldsymbol{q}, v)$ with worst-case variance $v \in [\check{v}, +\infty) \cap  [U_{\mathcal{M}}(\boldsymbol{q},\rho\cdot\theta^{U}),  U_{\mathcal{M}}(\boldsymbol{q},\rho\cdot\theta^{L})]$, as depicted by Line 2 in Alg. \ref{alg:pab_framework}.
\item Those instances of the key modules no longer need to guarantee the properties in Theorem \ref{theorem:arbitrage free}. Instead, they should guarantee the properties presented in Theorem \ref{thm:partial_ab} so that the pricing function $\Pi_{\mathcal{M}}$ is partially arbitrage free for $v\in[U_{\mathcal{M}}(\boldsymbol{q},\rho\cdot\theta^{U}),  U_{\mathcal{M}}(\boldsymbol{q},\rho\cdot\theta^{L})]$. Concretely, in our partially arbitrage-free trading framework, the \textit{PerturbMech} module should guarantee the second property and \textit{BudgetAlloc} should guarantee the others. 
\end{itemize}

\textbf{An option to support nonsubadditive compensation functions:} In our partially arbitrage-free trading framework, we also provide an option to achieve partial arbitrage freeness when some of data owners' compensation functions are not subadditive, e.g., superadditive. In our arbitrage-free trading framework, we assume that each data owner $u_i$'s compensation function $\mu_i$ is increasing and subadditive so that we can guarantee arbitrage freeness by Theorem \ref{theorem:arbitrage free}; also, Theorem \ref{thm:partial_ab} requires the same assumption. However, some data owners may prefer compensation functions that are not subadditive, as shown in the following example. 

\begin{exm}
\label{exm:super}
Given a superadditive compensation function $\exp(\epsilon^\tau_i)-1$ and a subadditive compensation function $\sqrt{\epsilon^\tau_i}$, to gain higher compensation, those data owners who expect themselves to sell a small privacy loss of $0.5$ for each query may prefer the subadditive compensation function because $\exp(0.5)-1 < \sqrt{0.5}$. In contrast, those who expect themselves to sell a large privacy loss of $1.0$ for each query may prefer the superadditive compensation function because $\exp(1.0)-1 >\sqrt{1.0}$. In addition, if only the subadditive compensation function is provided, data owners might not be encouraged enough to sell her privacy loss of a sufficient amount because the higher privacy loss corresponds to the lower unit compensation (i.e., $\frac{\mu_i(\epsilon_i)}{\epsilon_i}$).
\end{exm}

To support superadditive compensation functions in our partially arbitrage-free trading framework, we propose the following Theorem \ref{thm:partial_ab_sup} as a replacement of Theorem \ref{thm:partial_ab}. That is, in addition to subadditive compensation functions, if data owners are provided with nonsubadditive functions, the key modules should guarantee the properties in Theorem \ref{thm:partial_ab_sup} instead of those in Theorem \ref{thm:partial_ab}.

\begin{thm}
\label{thm:partial_ab_sup}
Given two constants $\theta^{U} \geq\theta^L > 0$, the pricing function $\pi=\Pi_{\mathcal{M}}(\mathbf{q}, v)=\boldsymbol{\mu}((U^{\mathbf{q}}_{\mathcal{M}})^{-1}(v))$ is partially arbitrage free for $v\in [U_{\mathcal{M}}(\boldsymbol{q},[f_1(\theta^{U}),...,f_n(\theta^{U})]),$ $U_{\mathcal{M}}(\boldsymbol{q},[f_1(\theta^{L}),...,f_n(\theta^{L})])]$ if the utility function $v=U_{\mathcal{M}}(\mathbf{q}, [\epsilon_1,...,\epsilon_n])$ satisfies:
\begin{enumerate}
\item the five properties in Theorem \ref{thm:partial_ab} are satisfied,
\item  $\forall i$, if $\mu_i$ is superadditive, it holds that 
\begin{equation*}
    U_{\mathcal{M}}(\boldsymbol{q},[f_1(\theta^{A}_i),...,f_n(\theta^{A}_i)])\leq \frac{U_{\mathcal{M}}(\boldsymbol{q},[f_1(\theta^{L}),...,f_n(\theta^{L})])}{2},
\end{equation*}
where $\theta^{A}_i= {f_i}^{-1}({\mu_i}^{-1}(2\mu_i(f_i(\theta^L))))$, and
\item  $\forall i$, if $\mu_i$ is superadditive, it holds that $\forall \theta \in [\theta^{L}, \theta^{U}]$, $v\cdot \frac{\partial^2 v}{\partial c_i^2}\leq 2(\frac{\partial v}{\partial c_i})^2$, where $c_i = \mu_i(\epsilon_i)$.
\end{enumerate}
\end{thm}

\begin{Rem}
Here, we explain the five properties in Theorem \ref{thm:partial_ab_sup}. The first property guarantees that the utility function corresponds to a partially arbitrage-free pricing function if all the compensation functions are subadditive. Then, the second and third properties additionally prevent the violation of partial arbitrage freeness due to those superadditive compensation functions.
Concretely, similar to the third property in Theorem \ref{thm:partial_ab}, the second property also requires an appropriate value of the endpoint $U_{\mathcal{M}}(\boldsymbol{q},[f_1(\theta^{L}),...,f_n(\theta^{L})])$ but takes the compensation functions into consideration. 
Then, the third property requires a rapid decreasing speed of the utility function with respect to each data owner's compensation $\mu_i(\epsilon_i)$, so that the price, which is equal to the sum of the compensations, decreases slowly as $v$ increases.
\end{Rem}

\section{Arbitrage-Free Trading Protocols}
\label{sec:ab_protocols}

In this section, we propose two arbitrage-free trading protocols by instantiating two key modules in Alg. \ref{alg:framework}: \textit{PerturbMech} and \textit{BudgetAlloc}.

\subsection{Baseline: UniformTrading}
We propose the UniformTrading protocol as our baseline, which utilizes the Laplace mechanism and Uniform as instances of \textit{PerturbMech} and \textit{BudgetAlloc}, respectively.  

\textbf{PerturbMech:} The Laplace mechanism is often used for implementing differential privacy and its variants, such as personalized differential privacy. In previous work \cite{li2014theory}, the broker trades query answers perturbed by the Laplace mechanism, which causes uniform privacy losses for all data owners. In this paper, even though the upper bound of privacy loss is controlled by each data owner, which results in personalized privacy loss bounds, we can still deploy the Laplace mechanism $Lap^{\boldsymbol{q}}_{\epsilon}$ to perturb query answers and achieve $[\epsilon,...,\epsilon]$-PDP (see Theorem \ref{thm:mechs}).
 
\textbf{BudgetAlloc:} However, according to Definition \ref{def:min_lap}, for each data owner $u_i$, since the Laplace mechanism $Lap^{\boldsymbol{q}}_{\epsilon}$ can only generate uniform privacy losses $[\epsilon,...,\epsilon]$, it is meaningless to allocate privacy budgets $\bar{\epsilon}_i > \min_j{\bar{\epsilon}_j}$. Hence, under the Laplace mechanism, we use Uniform (see Definition \ref{def:uniform}) to allocate uniform privacy budgets. In addition, we use a reservation ratio $r$ to reserve some privacy budgets for future queries.



\begin{Def}[Uniform]
\label{def:uniform}
Uniform allocates $[\bar{\epsilon}_1,...,\bar{\epsilon}_n]=[\min_i{\hat{\epsilon}_i},...,\min_i{\hat{\epsilon}_i}] * (1-r)$, where $r\in [0,1]$. 
\end{Def}

We can easily observe that under the Uniform algorithm, the pattern $\rho$  is equal to $[1,...,1]$, and the utility function $U_{Lap}(\boldsymbol{q}, \rho\cdot\theta)=2\cdot (\frac{\Delta\boldsymbol{q}}{\theta})^2$ satisfies the five properties in Theorem \ref{theorem:arbitrage free}. Thus, $\pi=\Pi_{Lap}(\boldsymbol{q}, v)$ is arbitrage free. However, UniformTrading is not ideal for data owners because it cannot make full use of personalized privacy loss bounds.

\begin{Coro}[of Theorem \ref{theorem:arbitrage free}]
Under UniformTrading, the pricing function $\pi=\Pi_{Lap}(\boldsymbol{q},v)$ is arbitrage free.
\end{Coro}
\subsection{Advanced: PersonalizedTrading}
\label{sec:personalizedTrading}

To make full use of data owners' personalized privacy loss bounds by personalized privacy losses, we propose the PersonalizedTrading protocol, which is a combination of the Sample mechanism as an instance of \textit{PerturbMech} and Patterning (with a preprocessing algorithm PatternSearch) as an instance of \textit{BudgetAlloc}.

\textbf{PerturbMech:} We leverage the Sample mechanism for module \textit{PerturbMech}. According to Theorem \ref{thm:mechs}, under the Sample mechanism, privacy losses can be personalized. 



\textbf{BudgetAlloc:} We propose an algorithm named Patterning (Alg. \ref{alg:patterning}) to compute privacy budgets that fit in a given $\rho$-pattern. Given a pattern $\rho$, Patterning finds the maximum privacy budgets that fit in the $\rho$-pattern within the privacy loss bounds and allocates a portion of them for the current query, i.e., $[\bar{\epsilon}_1,...,\bar{\epsilon}_n]=$Patterning($[\hat{\epsilon}_1,...,\hat{\epsilon}_n], \rho$)$=\rho\cdot\sup\{\theta|\forall i, \rho_i \cdot \theta\leq \hat{\epsilon}_i\} *(1-r)$. In addition, if the variable $exchange=True$, an optional function named PatternExchange changes the value of $\rho$ so that Patterning can dynamically compute higher privacy budgets when continuously trading multiple queries. 
We will introduce the PatternExchange algorithm after PatternSearch.

\begin{algorithm}[h]
\caption{Patterning}
\begin{algorithmic}[1]
\REQUIRE $[\hat{\epsilon}_1,...,\hat{\epsilon}_n]$, $\rho$ (can be precomputed by Alg. \ref{alg:pattern_search}), $exchange=False$, reservation ratio $r$
\ENSURE $[\bar{\epsilon}_1,...,\bar{\epsilon}_n]$
\IF{$exchange=True$}
	\STATE $\rho \gets$ PatternExchange($\rho$, $[\hat{\epsilon}_1,...,\hat{\epsilon}_n]$)
\ENDIF
\STATE $\theta \gets \max_i \hat{\epsilon}_i$
\FOR {$i=1$ \TO $n$}
	\IF {$\hat{\epsilon}_i < \theta * \rho_i$}
		\STATE $\theta \gets \hat{\epsilon}_i/\rho_i$
	\ENDIF
\ENDFOR
\RETURN $[\bar{\epsilon}_1,...,\bar{\epsilon}_n] \gets \theta*\rho * (1-r)$
\end{algorithmic}
\label{alg:patterning}
\end{algorithm}

\begin{algorithm}[h]
\caption{PatternSearch}
\begin{algorithmic}[1]
\REQUIRE $[\hat{\epsilon}_1,...,\hat{\epsilon}_n]$, an extremely small positive constant $\sigma$
\ENSURE $\rho$
\STATE $\rho, \rho^{start}, \rho^{end} \gets [\hat{\epsilon}_1,...,\hat{\epsilon}_n]/ \max_i \hat{\epsilon}_i$
\STATE For each pattern element $\rho^{end}_i \not = 1$, let $\rho^{end}_i = 0$
\WHILE {True}
	\STATE $\rho^{pre} \gets \rho$
	\IF {optimization constraints in Problem \ref{prom:arbitrage free} are satisfied,}
		\STATE $\rho^{end} \gets \rho$
		\STATE For each pattern element $\rho_i \not = 1$, let $\rho_i = (\rho_i + \rho^{start}_i) / 2$
	\ELSE
		\STATE $\rho^{start} \gets \rho$
		\STATE For each pattern element $\rho_i \not = 1$, let $\rho_i = (\rho_i+ \rho^{end}_i)/2$
	\ENDIF
	\IF {$\sum_i(\rho_i - \rho^{pre}_i)^2 < \sigma$}
		\RETURN $\rho^{end}$
	\ENDIF
\ENDWHILE
\end{algorithmic}
\label{alg:pattern_search}
\end{algorithm}

\textbf{PatternSearch (Alg. \ref{alg:pattern_search}):} Before we trade query answers, we should decide an optimal pattern $\rho$ as an input of Patterning to allocate privacy budgets. An ideal solution is to make the pattern $\rho$ of privacy budgets the same as the pattern $\rho^{bound}$ of the privacy loss bounds, i.e., $\rho=\rho^{bound}=[\hat{\epsilon}_1,...,\hat{\epsilon}_n]/ \max_i \hat{\epsilon}_i$, by which the broker can exhaust all the data owners' privacy loss bounds. Unfortunately, for some $\rho$, the utility function $v=U_{Sam}(\boldsymbol{q},\rho \cdot \theta)$ cannot satisfy the last two properties in Theorem \ref{theorem:arbitrage free} and thus may not correspond to an arbitrage-free pricing function. For example, given a $\rho$, $v=U_{Sam}(\boldsymbol{q},\rho\cdot\theta)$ might decrease too slowly as $\theta$ increases; in other words, $\theta$ and the price $\pi=\Pi_{Sam}(\boldsymbol{q},v)=\boldsymbol{\mu}({U^{\boldsymbol{q}}_{Sam}}^{-1}(v))$ might decrease too rapidly as $v$ increases, which allows arbitrage behaviors. In addition, because the utility function of the Sample mechanism is complex, i.e., $U_{Sam}(\boldsymbol{q},\rho\cdot\theta)=(\Delta\boldsymbol{q})^2 \cdot [\sum_i \Pr_i \cdot (1-\Pr_i) + \frac{2}{(\theta)^2}]$ where $\Pr_i=\frac{e^{\rho_i \cdot \theta}-1}{e^{\theta}-1}$, finding a pattern $\rho$ that results in an arbitrage-free pricing function is extremely difficult. In fact, we cannot express the inverse function ${U^{\boldsymbol{q}}_{Sam}}^{-1}$ in analytic form. However, given Theorem \ref{theorem:arbitrage free}, we can transform the problem of finding a $\rho$ resulting in an arbitrage-free pricing function under the Sample mechanism into the following optimization problem.
\begin{Prom}
\label{prom:arbitrage free}[Finding the Optimal Arbitrage-Free Pattern]
\begin{equation*}
\begin{aligned}
&\min\limits_{\rho} \sum_i (\rho_i\!- \rho^{bound}_i)^2,\\
&s.t.\text{ } \forall \theta\!>0, \text{ } \frac{\partial U_{Sam}}{\partial \theta}(\boldsymbol{q},\rho\cdot\theta)+\beta \leq 0, \text{ } and \\
& U_{Sam}(\boldsymbol{q},\rho\cdot\theta)\cdot\frac{\partial^2 U_{Sam}}{\partial (\theta)^2}(\boldsymbol{q},\rho\cdot\theta)-2(\frac{\partial U_{Sam}}{\partial \theta}(\boldsymbol{q},\rho\cdot\theta))^2 \!\leq 0.
\end{aligned}
\end{equation*}
\end{Prom}
By solving Problem \ref{prom:arbitrage free}, we can find a pattern $\rho$ that is closest to the pattern $\rho^{bound}$ by which we can make full use of the privacy loss bounds. 
The constraints in Problem \ref{prom:arbitrage free} are derived from the last two properties in Theorem \ref{theorem:arbitrage free}, where $\beta>0$ is an extremely small positive constant. We note that we skip the first three properties because they are always satisfied under the Sample mechanism and $\rho$-pattern. Since Problem \ref{prom:arbitrage free} is a nonconvex optimization problem that is generally NP-hard, we also propose a heuristic named PatternSearch (Alg. \ref{alg:pattern_search}) to solve it by binary search, based on the observation that for each element $\rho_i \not = 1$, the smaller the $\rho_i$ is, the closer the utility function $v=U_{Sam}(\boldsymbol{q}, \rho \cdot \theta)$ is to satisfying the optimization constraints.

\begin{Rem}
\label{rem:pattern_search}
We note that there must be such a $\rho$ found by PatternSearch that satisfies those constraints because in the worst case where each element $\rho_i$ is equal to $0$ or $1$, the utility function is $v=U_{Sam}(\boldsymbol{q}, \rho \cdot \theta)=2\cdot (\frac{\Delta \boldsymbol{q}}{\theta})^2$, which satisfies the five properties in Theorem \ref{theorem:arbitrage free}. In addition, by PatternSearch, the higher privacy loss bound $\hat{\epsilon}_i$ a data owner $u_i$ set, the higher pattern element $\rho_i$ she corresponds to in the pattern $\rho$; thus, data owners are encouraged to set a higher privacy loss bound $\hat{\epsilon}_i$ so that she can sell a higher privacy loss for each query when compared with other data owners.
\end{Rem}

\textbf{PatternExchange (Alg. \ref{alg:pattern_exchange}):} The PatternExchange algorithm serves as an optional function in Patterning. Thus far, the pattern $\rho$ should stay the same for every query. Then, to provide  higher privacy budgets for data buyers so that higher privacy losses might be consumed, PatternExchange dynamically changes the value of $\rho$ for each query. The following example provides insight into PatternExchange.

\begin{exm}
\label{exm:pattern_exchange}
Consider the case where the number of data owners is $n=3$, privacy loss bounds are $[\hat{\epsilon}_1,\hat{\epsilon}_2,\hat{\epsilon}_3]= [2,1.6,1.0]$, compensation functions are $\mu_1(\epsilon_1)=2\epsilon_1$, $\mu_2(\epsilon_2)=2\epsilon_2$ and $\mu_3(\epsilon_3)=3\epsilon_1$, and the pattern precomputed by PatternSearch is $\rho=[1,0.6,0.4]$. For the first query, a data buyer consumed privacy losses $[\epsilon_1,\epsilon_2,\epsilon_3]=[1.5,0.9,0.6]$ by pattern $\rho$. Then for the second query, given privacy budgets $[\hat{\epsilon}_1,\hat{\epsilon}_2,\hat{\epsilon}_3]=[0.5,0.7,0.4]$, a data buyer at most can consume privacy losses $[\epsilon_1,\epsilon_2,\epsilon_3]=[0.5,0.3,0.2]$ by pattern $\rho$; however, if we exchange the values of the first two elements of $\rho$ and derive a new pattern $\rho'=[0.6,1,0.4]$, then a data buyer at most can consume privacy losses $[\epsilon_1,\epsilon_2,\epsilon_3]=[0.42,0.7,0.28]$. Because $0.42+0.7+0.28>0.5+0.3+0.2$, higher privacy losses can be consumed if we use $\rho'$ for the second query.
\end{exm}

\begin{algorithm}[h]
\caption{PatternExchange}
\begin{algorithmic}[1]
\REQUIRE $\rho$, $[\hat{\epsilon}_1,...,\hat{\epsilon}_n]$
\ENSURE optimal equivalent pattern $\rho'$
\STATE $\rho' \gets \rho$
\STATE $groups \gets$  Group data owners by their compensation functions
\FOR {$group \in groups$}
	\STATE $sub\_pattern \gets$ Derive a vector of all the group members' pattern elements $\rho_i$ 
    \STATE Sort $sub\_pattern$ by the values of pattern elements.
	\STATE Sort $group$ by the values of the group members' (remaining) privacy loss bounds $\hat{\epsilon}_i$
    \FOR {$j\in range(length(group))$}
    	\STATE //A group member with a higher timeline privacy budget corresponds to a higher pattern element.
    	\STATE $group[j]$'s pattern element in $\rho' \gets sub\_pattern[j]$ 
    \ENDFOR
\ENDFOR
\RETURN $\rho'$
\end{algorithmic}
\label{alg:pattern_exchange}
\end{algorithm}

However, changing the values of the pattern $\rho$ precomputed by PatternSearch may result in a nonarbitrage-free pricing function because the pattern $\rho$ affects the properties of the utility function. Hence, the question arises of how to change $\rho$  in an arbitrage-free manner where the properties of the pricing function remain the same. We propose the following Theorem \ref{thm:equiv_pattern} to show that if two patterns $\rho, \rho'$ are equivalent (see Definition \ref{def:equiv_pattern}), then the corresponding pricing functions are equivalent. Based on Theorem \ref{thm:equiv_pattern}, by exchanging the values of two exchangeable elements $\rho_i, \rho_j$ of the pattern $\rho$, PatternExchange dynamically derives an equivalent pattern $\rho'$  of $\rho$ such that higher privacy budgets can be provided under the $\rho'$-pattern, i.e., Patterning($[\hat{\epsilon}_1,...,\hat{\epsilon}_n]$, $\rho'$, $False$)$\geq$Patterning($[\hat{\epsilon}_1,...,\hat{\epsilon}_n]$, $\rho$, $False$).

\begin{Def}[Equivalent Pattern]
\label{def:equiv_pattern}
Two elements $\rho_i, \rho_j$ of a pattern $\rho$ are exchangeable if $\forall \epsilon \geq 0, \mu_i(\epsilon) = \mu_j(\epsilon)$. Two patterns $\rho^a,\rho^b$ are equivalent (denoted as $\rho^a \equiv \rho^b$) if
\begin{enumerate}
\item  we can derive pattern $\rho^b$ by exchanging the values of two exchangeable elements $\rho^a_i, \rho^a_j$ of pattern $\rho^a$, or
\item  there exists a pattern $\rho^c$ such that $\rho^c \equiv \rho^a$ and $\rho^c \equiv \rho^b$.
\end{enumerate}
\end{Def}

\begin{thm}
\label{thm:equiv_pattern}
Let two patterns $\rho, \rho'$ be equivalent, i.e., $\rho \equiv \rho'$. The pricing functions $\pi=\Pi^{\rho}_{Sam}(\boldsymbol{q},v)=\boldsymbol{\mu}(\rho\cdot \theta)$ and $\pi=\Pi^{\rho'}_{Sam}(\boldsymbol{q},v)=\boldsymbol{\mu}(\rho'\cdot \theta)$ are equivalent, i.e., $\forall Q=(\boldsymbol{q}, v), \Pi^{\rho}_{Sam}(\boldsymbol{q},v) = \Pi^{\rho'}_{Sam}(\boldsymbol{q},v)$. 
\end{thm}

\begin{Rem}
Intuitively, because two data owners with the same compensation function are identical from the perspective of pricing, their pattern elements are exchangeable. That is, if we exchange any two exchangeable pattern elements in pattern $\rho$, the derived pattern $\rho'$ will not cause a change in the pricing function. Therefore, even if PatternExchange dynamically changes the pattern for budget allocation, the pricing function stays arbitrage free.
\end{Rem}

In conclusion, PatternSearch computes a pattern $\rho$ that corresponds to a utility function satisfying the five properties in Theorem \ref{theorem:arbitrage free} and an arbitrage-free pricing function. PatternExchange guarantees that even if it changes the pattern $\rho$ dynamically, the pricing function stays the same. Thus, the following corollary is true, and PersonalizedTrading is an arbitrage-free trading protocol.

\begin{Coro}[of Theorem \ref{theorem:arbitrage free} and Theorem \ref{thm:equiv_pattern}]
Under the PersonalizedTrading protocol, given a pattern $\rho$ found by PatternSearch, the pricing function $\pi=\Pi_{Sam}(\boldsymbol{q},v)$ is arbitrage free.
\end{Coro}

\section{Partially Arbitrage-Free Trading Protocols}
\label{sec:pab_protocols}
Based on our partially arbitrage-free trading framework, we also propose two partially arbitrage-free trading protocols by instantiating the key modules, i.e., the UniformTradingPlus and PersonalizedTradingPlus protocols.

\subsection{UniformTradingPlus}
The UniformTradingPlus protocol is a partially arbitrage-free version of UniformTrading since it also employs the Laplace mechanism and the Uniform algorithm. The difference is that while UniformTrading requires subadditive compensation functions to guarantee arbitrage freeness, UniformTradingPlus supports some compensation functions that are not subadditive and guarantees partial arbitrage freeness. By Theorem \ref{thm:partial_ab_sup}, we propose the following Corollary \ref{prop:lap_partial} to show that the pricing function $\Pi_{Lap}$ under UniformTradingPlus can be partially arbitrage free.

\begin{Coro}[of Theorem \ref{thm:partial_ab_sup}]
\label{prop:lap_partial}
The pricing function $\pi = \Pi_{Lap}(\boldsymbol{q}, v) = \sum_i \mu_i(\Delta\boldsymbol{q}\cdot\sqrt{\frac{2}{v}})$ is partially arbitrage free for $v \in [2(\frac{\Delta\boldsymbol{q}}{\theta^{U}})^2, 2(\frac{\Delta\boldsymbol{q}}{\theta^L})^2]$ if the following two properties are satisfied:
\begin{enumerate}
\item  $\forall \mu_i, \mu_i(\sqrt{2}\theta^L) \leq 2\mu_i(\theta^L)$, and
\item  $\forall \mu_i, \forall \theta \in [\theta^L, \theta^{U}], \theta\leq \frac{\mu_i'(\theta)}{\mu_i''(\theta)}$
\end{enumerate}
\end{Coro}

\begin{proofnoof}[Corollary \ref{prop:lap_partial}]
If the properties are satisfied, the utility function $U_{Lap}$  satisfies the properties in Theorem \ref{thm:partial_ab_sup}.
\end{proofnoof}

\textbf{Partially arbitrage-free range selection:} 
Under UniformTradingPlus, given data owners' compensation functions, there might be multiple intervals of the worst-case variance $v$ that can serve as the partially arbitrage-free range. We can first let $\theta^L=\inf\{\theta|\forall \mu_i, \theta\leq \frac{\mu_i'(\theta)}{\mu_i''(\theta)}, \mu_i(\sqrt{2}\theta^L) \leq 2\mu_i(\theta), \theta\geq 0 \}$ and then find $\theta^{U}=\arg\max_{\theta^{U}\geq \theta^L}\theta^{U} - \theta^L$ with the constraint that the second property in Corollary \ref{prop:lap_partial} is satisfied. Finally, we can select the interval $[2(\frac{\Delta\boldsymbol{q}}{\theta^{U}})^2, 2(\frac{\Delta\boldsymbol{q}}{\theta^L})^2]$ as the partially arbitrage-free range.

\begin{Rem}
Because UniformTrading enforces a uniform privacy loss $\epsilon$ for all data owners, we cannot make full use of privacy loss bounds $[\hat{\epsilon}_1,...,\hat{\epsilon}_n]$ when there exists a data owner $u_j$'s privacy loss bound that is much lower than the others. To incentivize data owners to set a higher privacy loss bound $\hat{\epsilon}_i$, we propose the UniformTradingPlus protocol, by which data owners can be provided with superadditive compensation functions for selection.
\end{Rem}

\subsection{PersonalizedTradingPlus}
The PersonalizedTradingPlus protocol is a partially arbitrage-free version of PersonalizedTrading, but it can make higher use of data owners' personalized privacy loss bounds. Concretely, under PersonalizedTradingPlus, PatternSearch solves the following optimization problem to search a pattern $\rho$ such that the pricing function $\Pi_{Sam}$ is partially arbitrage free; that is, it differs from PersonalizedTrading in that Line 5 in PatternSearch is replaced with "\textbf{if} optimization constraints in Problem \ref{prom:partial_ab} are satisfied, \textbf{then}". 

\begin{Prom}
\label{prom:partial_ab}[Finding the Optimal Partially Arbitrage-Free Pattern]
\begin{equation*}
\begin{aligned}
&\min\limits_{\rho} \sum_i (\rho_i\!- \rho^{bound}_i)^2, \\
&s.t.\text{ } \forall \theta \geq \theta^L, \theta \leq \theta^U -\theta^L, U_{Sam}(\boldsymbol{q},\rho\cdot (\theta + \theta^{L}))\\ 
&\leq 1/(\frac{1}{U_{Sam}(\boldsymbol{q},\rho \cdot \theta^{L})} + \frac{1}{U_{Sam}(\boldsymbol{q},\rho \cdot \theta)}),\\
&\forall \theta > 0, \frac{\partial U_{Sam}}{\partial \theta}+\beta \leq 0, \text{ and } \\
&\forall \theta\in [\theta^L,\theta^{U}],  U_{Sam}(\boldsymbol{q},\rho\cdot\theta)\cdot\frac{\partial^2 U_{Sam}}{\partial (\theta)^2}-2(\frac{\partial U_{Sam}}{\partial \theta})^2 \leq 0.
\end{aligned}
\end{equation*}
\end{Prom}
where the constraints are derived from the last three properties in Theorem \ref{thm:partial_ab}, so that the pattern $\rho$ found by solving this problem corresponds to a pricing function $\Pi_{Sam}$ that is partially arbitrage free for  $v \in [U_{Sam}(\boldsymbol{q},\rho\cdot\theta^{U}),  U_{Sam}(\boldsymbol{q},\rho\cdot\theta^{L})]$. Because the constraints in Problem \ref{prom:partial_ab} are relaxed from the constraints in Problem \ref{prom:arbitrage free}, the pattern searched can be closer to the pattern $\rho^{bound}$, by which we can promote the use of the privacy loss bounds.

\textbf{Partially arbitrage-free range selection:} Ideally, the end point $U_{Sam}(\boldsymbol{q},\rho\cdot\theta^{U})$ should be no higher than $\check{v}$ so that we can answer those queries $Q=(\boldsymbol{q}, v)$ with $v \in [\check{v},  U_{Sam}(\boldsymbol{q},\rho\cdot\theta^{L})]$. To realize this, we can set $\theta^U=\sup\{\theta|\forall i, \rho_i \cdot \theta\leq \hat{\epsilon}_i\}$ in Problem \ref{prom:partial_ab} so that the privacy budgets $\rho\cdot \theta^{U}$ are maximized given a pattern $\rho$.

\subsection{Benefits of Partial Arbitrage Freeness}
Here, we highlight and discuss the benefits of adopting partial arbitrage freeness. We draw the importance of partial arbitrage freeness because other works focusing on arbitrage-free pricing can also benefit from it. 


\textbf{Relaxing the constraint on the compensation function:}
By partial arbitrage freeness, our marketplace can provide more types of compensation functions for selection. To encourage data owners to participate in personal data trading, the marketplace proposed in \cite{li2014theory} and our work compensate each data owner for her privacy loss. Because of the constraint of arbitrage freeness, Li et al. \cite{li2014theory} required that each compensation function should be a subadditive and nondecreasing function of privacy loss. However, as depicted in Example \ref{exm:super}, a data owner who expects herself to sell a large privacy loss may prefer a superadditive compensation function; if only subadditive compensation functions are provided, she might be unwilling to join in data trading. According to our Theorem \ref{thm:partial_ab_sup}, even if each compensation function is not subadditive, e.g., superadditive, the pricing function can still be partially arbitrage free. Therefore, partial arbitrage freeness can enable more types of compensation functions, which is an important incentive to data owners.

\textbf{Relaxing the constraint on the perturbation mechanism:}
Partial arbitrage freeness also allows more perturbation mechanisms to work for personal data pricing.
As shown in Theorem \ref{theorem:arbitrage free}, the selection of the perturbation mechanism plays a key role in the design of the pricing function. To achieve arbitrage freeness, the pricing methods in \cite{li2014theory} center around the particular Laplace mechanism. 
However, there are other perturbation mechanisms that satisfy differential privacy, such as the Exponential mechanism. As depicted in Example \ref{exm:pab-exp}, it might fail to guarantee arbitrage freeness with the Exponential mechanism, but the mechanism can work under partial arbitrage freeness. 
Moreover, if we consider the Sample mechanism with a certain pattern as a restricted perturbation mechanism, because of the relaxation, more patterns can be selected to restrict the Sample mechanism, and thus, more use of privacy loss bounds might be promoted. 
Therefore, partial arbitrage freeness can enable more perturbation mechanisms with better performances than the Laplace mechanism.

\begin{figure}[ht]
    \centering
    \includegraphics[scale=0.33]{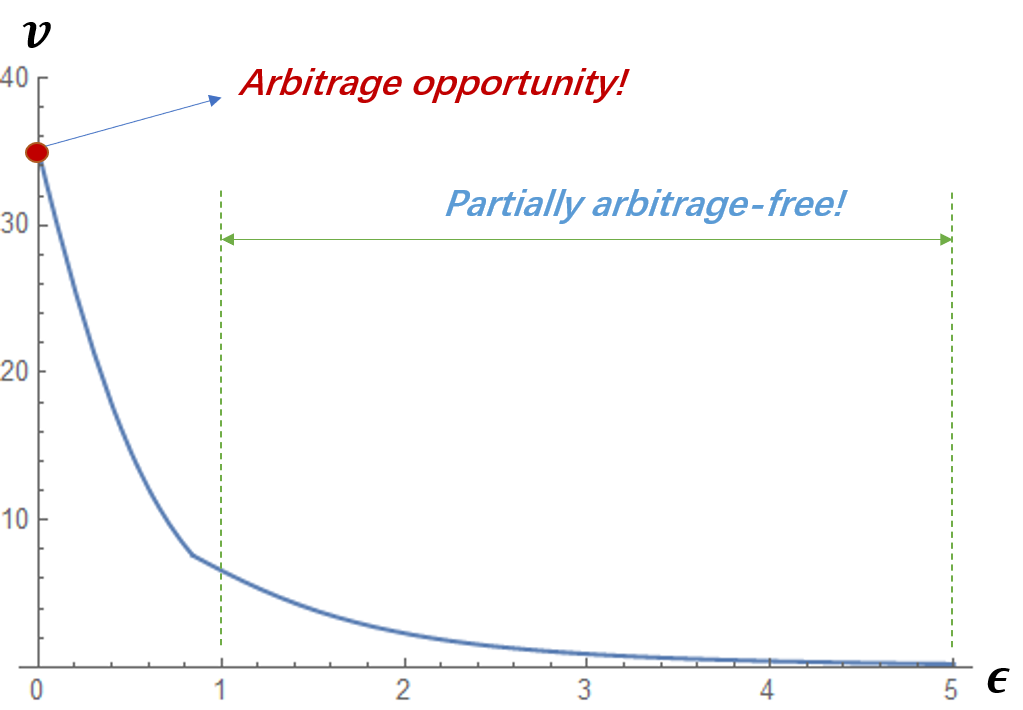}
    \caption{The utility function of the Exponential mechanism.}
    \label{fig:expo}
\end{figure}

\begin{exm}
\label{exm:pab-exp}
For simplicity, consider $n=10$ data owners, a linear counting query $\boldsymbol{q}$ with at least one zero-value element and others of one-value, a database $D$, a privacy loss $\epsilon$ for all data owners, and subadditive compensation functions. The Exponential mechanism $Expo^{\boldsymbol{q}}_\epsilon$ outputs each possible answer $o \in  \{0, 1,...,10\}$ with the probability 
\begin{equation*}
    Pr[Expo^{\boldsymbol{q}}_\epsilon= o]= \frac{\exp(-\epsilon|o-\boldsymbol{q}(D)|/2)}{\sum_r \exp(-\epsilon|r-\boldsymbol{q}(D)|/2)}.
\end{equation*}
Figure \ref{fig:expo} draws the utility function of $Expo^{\boldsymbol{q}}_\epsilon$, i.e., 
\begin{equation*}
    v=U_{Expo}(\boldsymbol{q}, [\epsilon,...,\epsilon])= \max_{y} \sum_{o=0}^{n} \frac{\exp(-\frac{\epsilon|o-y|}{2})(o-y)^2}{\sum_{r=0}^{n} \exp(-\frac{\epsilon|r-y|}{2})}.
\end{equation*}
Therefore, a data buyer can request a perturbed query answer with $v=35$ for free, which obviously violates arbitrage freeness. However, for $\epsilon \in [1.0, 5.0]$, the utility function $v=U_{Expo}(\boldsymbol{q}, [\epsilon,...,\epsilon])$ satisfies the properties in Theorem \ref{thm:partial_ab}, which means that the corresponding pricing function is partially arbitrage free for $v \in [U_{Expo}(\boldsymbol{q}, [5,...,5]), U_{Expo}(\boldsymbol{q}, [1,...,1])]$ according to Theorem \ref{thm:partial_ab}.
\end{exm}
\section{Experiments}
\label{sec:expr}


\begin{figure*}[t]
\centering
\begin{minipage}[t]{0.5\linewidth}
\subfigure[Minimum privacy loss bound.]{
\includegraphics[scale=0.18]{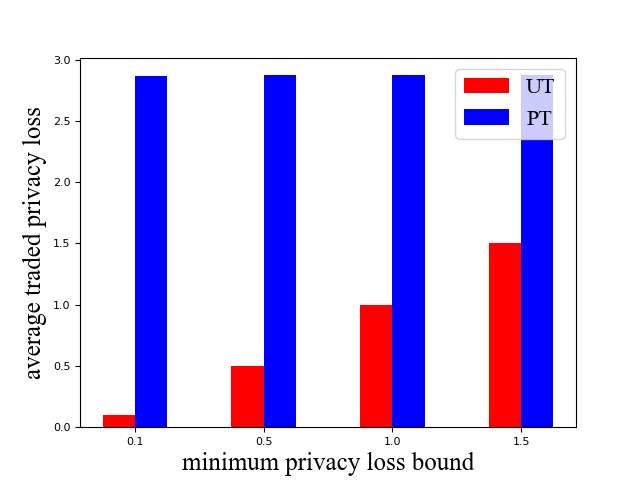}
\label{fig:rq1_min}
}
\subfigure[Maximum privacy loss bound.]{
\includegraphics[scale=0.045]{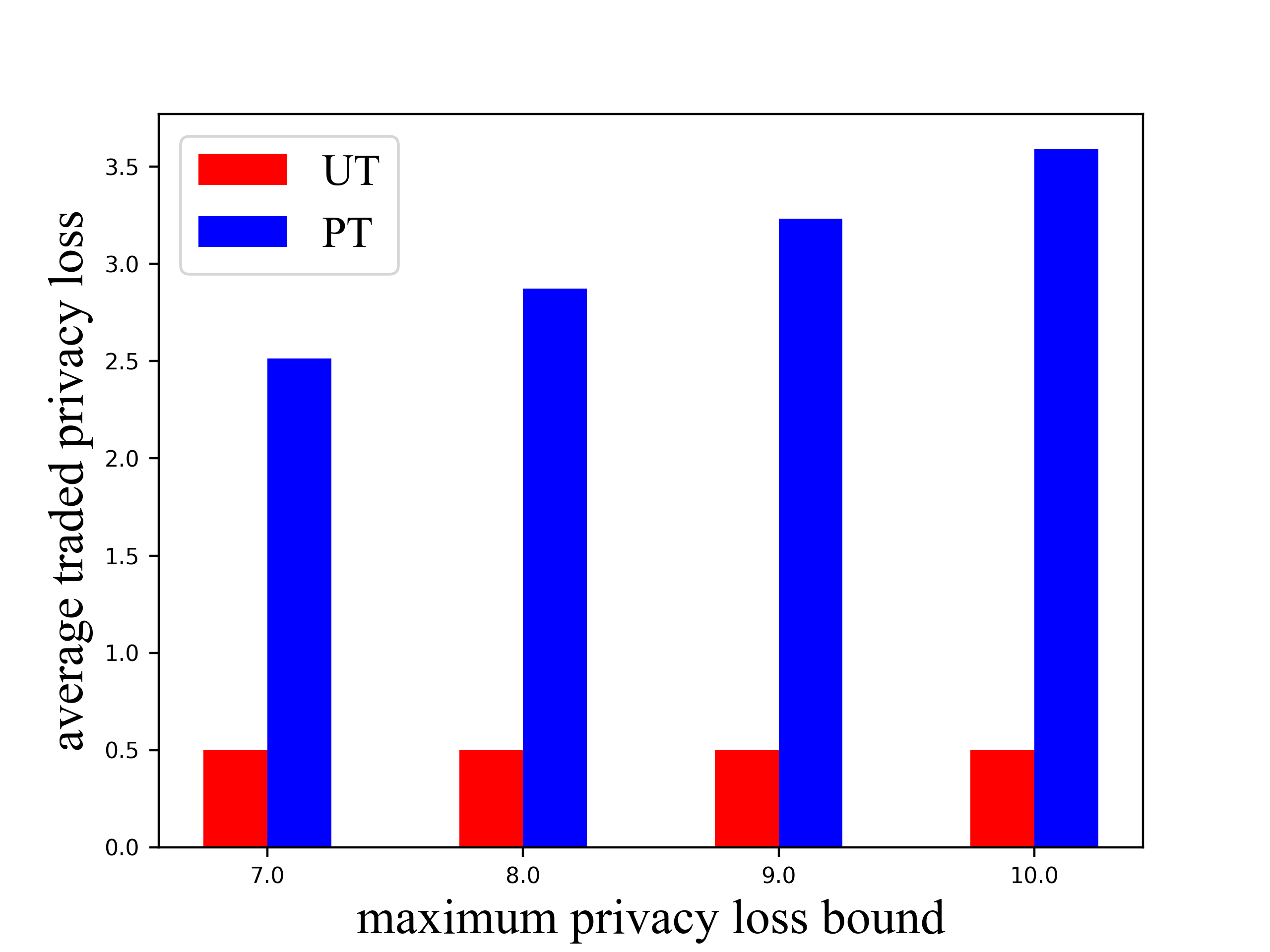}
\label{fig:rq1_max}
}
\caption{The effect of personalized privacy losses (RQ1).}
\label{fig:rq1}
\end{minipage}%
\begin{minipage}[t]{0.5\linewidth}
\subfigure[Partially arbitrage-free range.]{
\includegraphics[scale=0.18]{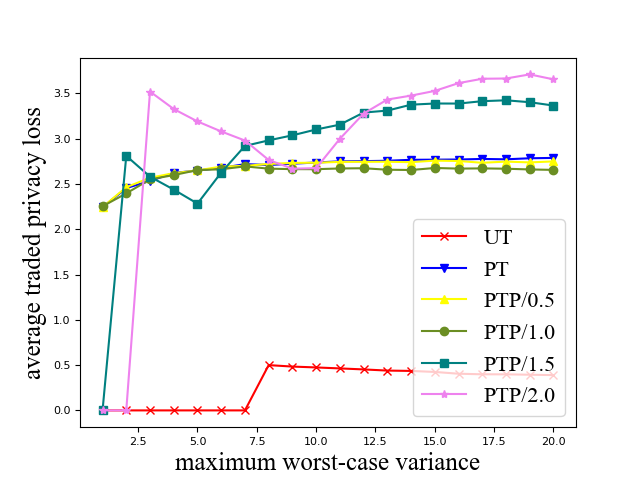}
\label{fig:rq2_range}
}
\subfigure[Reservation rate.]{
\includegraphics[scale=0.18]{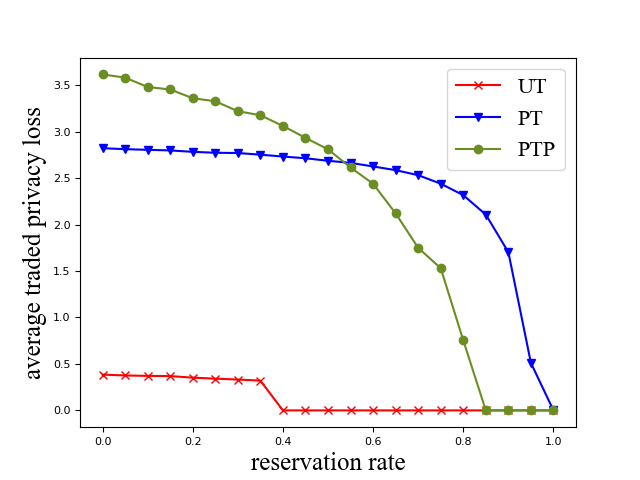}
\label{fig:rq2_res}
}
\caption{The effect of partial arbitrage freeness (RQ2).}
\label{fig:rq2}
\end{minipage}%
\centering
\end{figure*}

\begin{figure*}[t]
\centering
\begin{minipage}[t]{0.5\linewidth}
\subfigure[PersonalizedTrading]{
\includegraphics[scale=0.045]{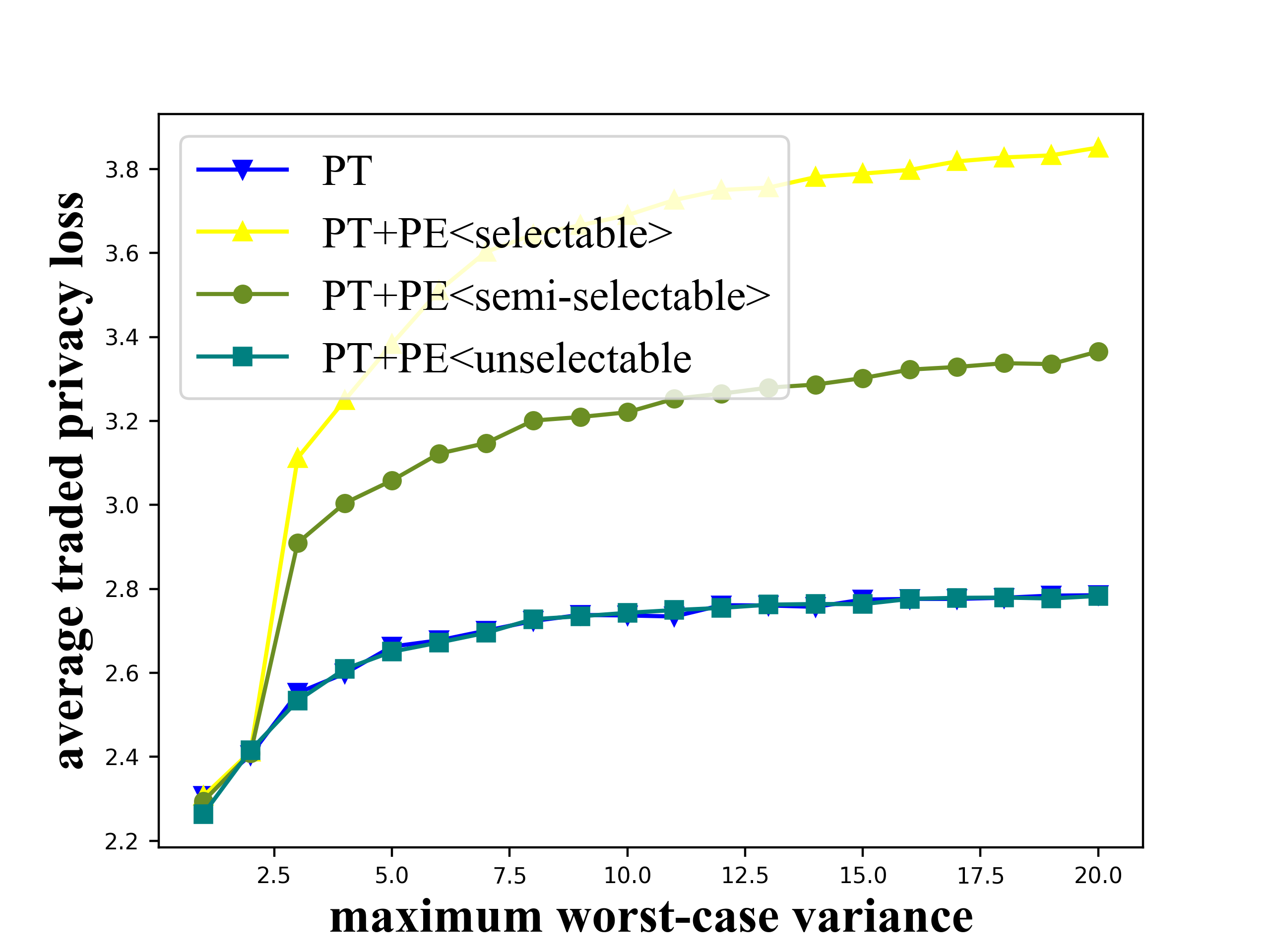}
\label{fig:rq3_pt}
}
\subfigure[PersonalizedTradingPlus]{
\includegraphics[scale=0.045]{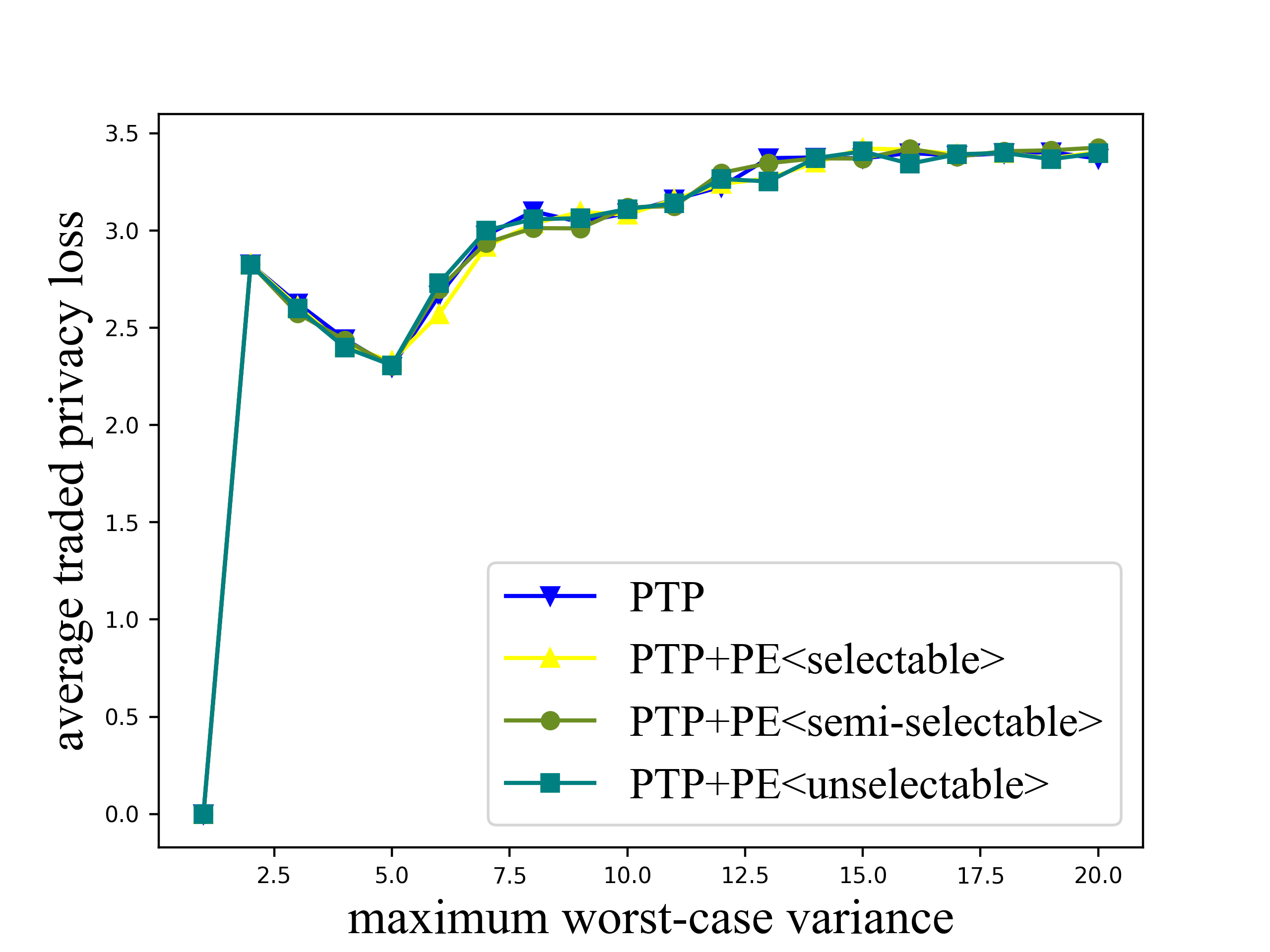}
\label{fig:rq3_ptp}
}
\caption{The effect of PatternExchange (RQ3).}
\label{fig:rq3}
\end{minipage}%
\begin{minipage}[t]{0.5\linewidth}
\subfigure[Arbitrage freeness (RQ4).]{
\includegraphics[scale=0.18]{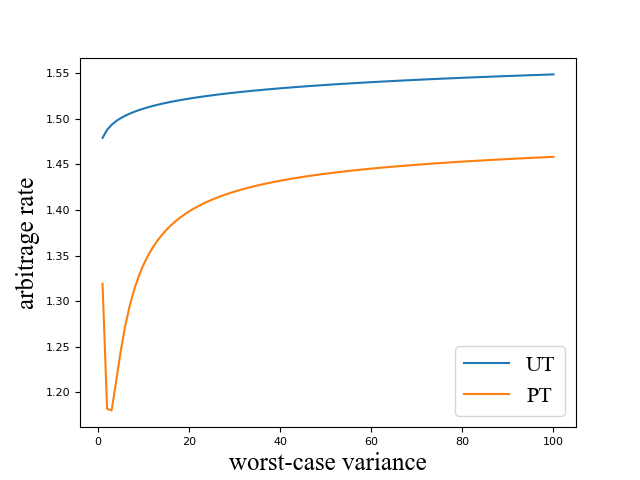}
\label{fig:rq4}
}
\subfigure[Partial arbitrage freeness (RQ5).]{
\includegraphics[scale=0.18]{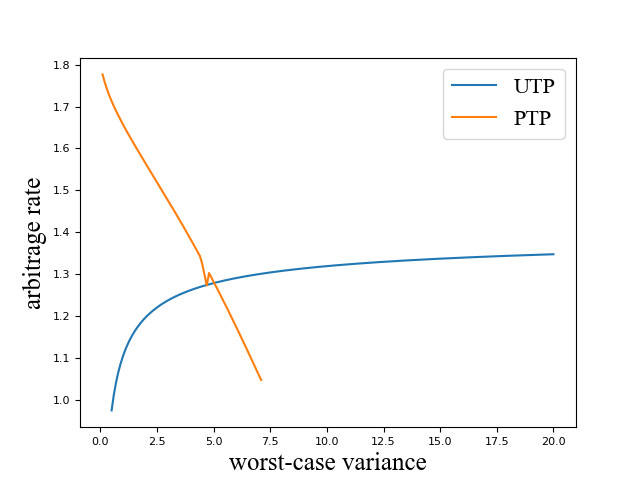}
\label{fig:rq5}
}
\caption{Simulating arbitrage attacks (RQ4 and RQ5).}
\label{fig:rq45}
\end{minipage}%
\centering
\end{figure*}


In this section, we simulate transactions in our marketplace to test the performance of the proposed trading protocols. 
\subsection{Experiment Setup}

\textbf{Research questions:} 
We conduct several parts of the experiments to answer some research questions. 
Concretely, for data owners, we evaluate the performance of our trading protocols in terms of making full use of their privacy loss bounds and answer the following questions:
\begin{itemize}
    \item RQ1: How does the perturbation mechanism with personalized privacy losses promote the use of privacy loss bounds? What's the effect of data owners' privacy loss bounds?
    \item RQ2: How does partial arbitrage freeness promote the use of privacy loss bounds? What are the effects of the partially arbitrage-free range and the reservation rate?
    \item RQ3: How does PatternExchange promote the use of privacy loss bounds? What is the effect of the distribution of data owners' compensation functions? 
\end{itemize}
Then, for the broker, we simulate arbitrage behaviors under each trading protocol to answer the following questions.
\begin{itemize}
    \item RQ4: Do the UniformTrading and PersonalizedTrading protocols satisfy arbitrage freeness?
    \item RQ5: Do UniformTradingPlus and PersonalizedTradingPlus achieve partial arbitrage freeness? Can UniformTradingPlus support superadditive compensation functions?
\end{itemize}
 
\textbf{Dataset:} 
We generate $n=200$ data owners' data points, and the value of each data point is randomly selected from $d=20$ values.
Then, according to a user survey on privacy preference \cite{nget2017balance}, we randomly divide those data owners into four groups: \textit{conservative}, consisting of $16$ percent of data owners with privacy loss bounds $\hat{\epsilon}_i = 0.5$; \textit{hesitant}, consisting of $16$ percent with $\hat{\epsilon}_i = 2.0$; \textit{ordinary}, consisting of $33$ percent with $\hat{\epsilon}_i = 4.0$; \textit{liberal}, consisting of $34$ percent with $\hat{\epsilon}_i = 8.0$. However, we may vary the privacy loss bound for the conservative group (i.e., the minimum privacy loss bound) and the bound for the liberal group (i.e., the maximum privacy loss bound) to answer RQ1.

\textbf{Compensation schemes:}
Basically, we provide four compensation functions for selection, i.e., a subadditive function $\mu^{B}(\epsilon_i)=2\sqrt{\epsilon_i}$, a linear function $\mu^{L}(\epsilon_i)=2\epsilon_i$, and two compound functions $\mu^{C1}(\epsilon_i)=\epsilon_i+\sqrt{\epsilon_i}$ and $\mu^{C2}(\epsilon_i)=1.5 \epsilon_i+ 0.5 \sqrt{\epsilon_i}$.
We also provide the following three compensation schemes:
\begin{itemize}
\item \textit{Selectable}: Each data owner randomly selects a compensation function from the above four functions.
\item \textit{Semiselectable}: Those conservative data owners select $\mu^{B}$, those hesitant ones randomly choose one from $\mu^{B}$ and $\mu^{C1}$, those ordinary ones randomly select one from $\mu^{C1}$ and $\mu^L$, and the others select $\mu^L$. 
\item \textit{Unselectable}: Those conservative data owners select $\mu^{B}$, those hesitant ones select $\mu^{C1}$, those ordinary ones select $\mu^{C2}$, and the others select $\mu^L$. 
\end{itemize}
By default, we employ the semiselectable scheme; when evaluating PatternExchange, we test all three schemes to answer RQ3. In addition, to answer RQ5, we may also let all the data owners select a superadditive compensation function.


\textbf{Parameters:}
For simplicity and fair comparison, we assume that all the linear queries in our experiments are of global sensitivity $\Delta\boldsymbol{q}=1$. 
For budget allocation, we set the reservation rate $r=0.2$ by default to reserve a part of the remaining privacy loss bounds for future queries, which also achieves satisfactory performance in terms of making full use of privacy loss bounds.
For the partially arbitrage-free range of the PersonalizedTradingPlus protocol, we set $\theta^L=1.5$ (by default) that promises a high use of privacy loss bounds and $\theta^{U}=10.0$ to cover all the values of the maximum privacy loss bound we test in the experiments.

\textbf{Evaluation methods:} We present the methods and metrics for evaluating the performance of each trading protocol in terms of achieving our design goals as follows. 
\begin{itemize}[leftmargin=*]
\item To make full use of privacy loss bounds, we evaluate the performance by the privacy loss consumed per data owner.
Concretely, for each trading protocol, we let data buyers request $100$ queries;  for each query, a buyer randomly selects the worst-case variance $v\in [\check{v}, V]$, where the parameter $V$ ($=100$ by default) denotes the \textit{maximum worst-case variance} accepted by data buyers. Then, we calculate the privacy losses $\epsilon_1,...,\epsilon_n$ from those queries and compute the average traded privacy loss $\frac{\sum_{i}\epsilon_i}{n}$ for evaluation. Due to the randomized queries, we run $100$ rounds of evaluation to obtain the expectation result.
\item To verify the arbitrage freeness or parital arbitrage freeness of a trading protocol, we simulate arbitrage attacks against its pricing function. 
Concretely, to arbitrage over a query $(\boldsymbol{q}, v)$, a savvy data buyer can conduct arbitrage attacks by purchasing $m \geq 2$ cheaper queries $(\boldsymbol{q}, m\cdot v)$ and aggregating them into the query $(\boldsymbol{q}, v)$. 
Given a pricing function $\pi=\Pi_{\mathcal{M}}(\boldsymbol{q}, v)$, if the cost of purchasing those $m$ queries is lower than the market price of the aggregated query, i.e., $m \cdot \Pi_{\mathcal{M}}(\boldsymbol{q}, m \cdot v) < \Pi_{\mathcal{M}}(\boldsymbol{q}, v)$, the buyer can do arbitrage successfully. 
In our experiments, we test $m\in [2, 10]$ and evaluate the arbitrage freeness or partial arbitrage freeness by the ratio between the minimum arbitrage cost and the market price, i.e., arbitrage rate $\frac{\min_{m \in [2,10]} m\cdot \Pi_{\mathcal{M}}(\boldsymbol{q}, m\cdot v)}{\Pi_{\mathcal{M}}(\boldsymbol{q}, v)}$. 
\end{itemize}


\subsection{Experiments for Data Owners}

For data owners, we verify the effectiveness of different trading protocols in terms of making full use of privacy loss bounds, and answer RQ1, RQ2 and RQ3.

\textbf{Effect of personalized privacy losses (RQ1):} 
As depicted in Figure \ref{fig:rq1}, we can observe that the average traded privacy loss under the PersonalizedTrading protocol (PT) is always higher than that under the UniformTrading protocol (UT) because the privacy losses can be personalized under PT but should be uniform under UT. 
Then, we also vary the privacy loss bounds of those data owners in the conservative and liberal groups, i.e., the minimum privacy loss bound and the maximum privacy loss bound. 
We can see that as the minimum privacy loss bound increases, the average traded privacy loss also increases under UT but stays almost the same under PT, which demonstrates that UT is more sensitive to the privacy loss bound of the most conservative data owners because the privacy losses have to be uniform and covered by the minimum privacy loss bound under UT.
With the maximum privacy loss bound increasing, the result is the opposite because PT always consumes the highest privacy losses of those liberal data owners by a pattern. 

\textbf{Effect of partial arbitrage freeness (RQ2):}
As shown in Figure \ref{fig:rq2_range}, in most cases of the maximum worst-case variance $V$, the partially arbitrage-free PersonalizedTradingPlus protocol (PTP) can consume a higher average traded privacy loss than PT, which demonstrates the benefit of partial arbitrage freeness, i.e., promoting the use of privacy loss bounds.
Concretely, we vary the value of the parameter $\theta^L$ to test PTP with different partially arbitrage-free ranges (denoted as PTP/$\theta^L$ in Figure \ref{fig:rq2}). We find that PTP performs slightly better than PT when $\theta^L\in\{0.5, 1.0\}$ but significantly better when $\theta^L \in \{1.5, 2.0\}$. 
That is because, under the Sample mechanism, the fifth property in Theorem \ref{theorem:arbitrage free} is hard to satisfy when $\theta \in (1.0, 1.5)$; 
if we turn to partial arbitrage freeness and do not guarantee the property for $\theta \in (0, 1.5)$, the pattern $\rho$ found by solving \ref{prom:partial_ab} can be closer to the pattern $\rho^{bound}$ of the privacy loss bounds than by solving Problem \ref{prom:arbitrage free}. 
However, as the reservation rate $r$ increases (see Figure \ref{fig:rq2_res}), the performance of PTP declines more steeply than PT. 
This happens because when $r$ is high, the privacy budgets for a query are also highly limited, which may result in an inaccurate query answer that cannot be traded under PTP and no privacy loss.

\textbf{Effect of PatternExchange (RQ3):}
As illustrated in Figure \ref{fig:rq3_pt}, if we run the function PatternExchange (PE), we can further significantly improve the performance of PT (i.e., PT+PE) because PE can flexibly change the pattern $\rho$ into an equivalent pattern $\rho'$, which is closer to the pattern of the remaining privacy loss bounds. We also test PE under different compensation schemes (denoted as PE$<$compensation scheme$>$ in Figure \ref{fig:rq3}). 
It can be observed that PE$<$selectable$>$ can utilize more privacy loss than the others because it can exchange more pattern elements of those data owners from different groups who share the same compensation functions .
However, Figure \ref{fig:rq3_ptp} shows that PE is not helpful for PTP because the pattern used in PTP is close enough to the pattern of the privacy loss bounds.

\subsection{Experiments for the Broker}
We simulate arbitrage attacks over our trading protocols to verify their arbitrage freeness or partial arbitrage freeness. 

\textbf{Arbitrage freeness (RQ4):} 
Finally, we simulate arbitrage attacks to verify the arbitrage freeness of UniformTrading (UT) and PersonalizedTrading (PT). 
As depicted in Figure \ref{fig:rq4}, under either UT or PT, for each worst-case variance $v$, the arbitrage cost $m\cdot \Pi_{\mathcal{M}}(\boldsymbol{q},m
\cdot v) $ is always higher than the market price $\Pi_{\mathcal{M}}(\boldsymbol{q},v) $ offered by the broker, which means the pricing function $\Pi_{\mathcal{M}}$ is arbitrage free.

\textbf{Partial arbitrage freeness (RQ5):}
We also conduct arbitrage attacks over UniformTradingPlus (UTP) and PersonalizedTradingPlus (PTP). For UniformTradingPlus, we let all the data owners select the superadditive compensation function $\mu^{S}(\epsilon_i)=\exp(\epsilon_i)-1$ to test the benefit of partial arbitrage freeness, i.e., supporting more types of compensation functions than arbitrage freeness. According to Corollary $\ref{prop:lap_partial}$, UTP should be partially arbitrage free for $v\geq 2$, which can be verified by Figure \ref{fig:rq5}. For PTP, since we set $\theta^L = 1.5$ and $\theta^U = 10.0$, the partially arbitrage free range of $v$ should be at least from $U_{Sam}(\boldsymbol{q}, \rho\cdot 10.0) \approx 0.11$ to $U_{Sam}(\boldsymbol{q}, \rho\cdot 1.5) \approx 14.22$. We can observe from Figure \ref{fig:rq5} that the arbitrage cost is higher than the market price for $v\in [0.1, 7.1]$; for $v>7.1$, arbitrage attacks cannot happen because the buyer is not allowed to purchase a query $(\boldsymbol{q}, m\cdot v)$ for any $m \geq 2$. Therefore, the partial arbitrage freeness of PTP can also be verified.

\section{Related Work}
\label{sec:related_work}
Balazinska et al. \cite{balazinska2011data} guided the tendency of the research on \textit{Data Market} for the database community and introduced the challenge of preventing arbitrage opportunities in \textit{query-based data pricing} models.
Inspired by the vision paper \cite{balazinska2011data}, Koutris et al. \cite{koutris2012query, koutris2015query} formalized the first theoretical framework and proposed the notion of \textit{arbitrage freeness} for pricing full relational queries, while almost at the same time Li et al. \cite{li2012pricing} studied the arbitrage issue for linear aggregate queries for the first time.
Then, both Lin et al. \cite{lin2014arbitrage} and Deep et al. \cite{deep2017design} provided mathematical conditions for avoiding arbitrage in pricing query bundles but considered different pricing schemes.
Miao et al. \cite{miao2020towards} also proposed a practical arbitrage-free pricing mechanism for querying incomplete data.
In addition, two concrete pricing systems QueryMarket \cite{koutris2013toward} and QIRANA \cite{deep2017qirana} have been developed based on the theoretical frameworks proposed by Koutris et al. \cite{koutris2012query, koutris2015query} and Deep et al. \cite{deep2017design}, respectively.
Chawla et al. \cite{chawla2019revenue} further addressed the problem of how to maximize the broker's revenue through query pricing with the guarantee of arbitrage freeness.
Recently, Fernande et al. \cite{fernandez2020data} provided their vision of a practical data market platform that can discover and integrate diverging data sources for data buyers; they also claimed the need of ensuring arbitrage-free pricing in their setting, i.e., preventing data buyers from obtaining the same data through a different and cheaper integration of data sources.
In this paper, to relax the challenging guarantee of arbitrage freeness, for the first time, we propose a novel and practical concept, i.e., partial arbitrage freeness, which means ensuring arbitrage-free pricing only for a certain part of queries and can be widely adopted in query-based pricing frameworks to expand the space of market design.

In addition, the above works did not consider the privacy leakage from trading queries. 
To protect privacy, some works, such as \cite{zhang2020privacy,niu2017trading}, employed cryptographic techniques to encrypt sensitive information when setting up a data marketplace. 
Other works, such as \cite{ghosh2011selling, ghosh2015selling,li2013theory,li2014theory,jung2019privacy,chen2019towards,liu2021dealer}, used differential privacy (DP \cite{dwork2006calibrating}) to add noise to the raw data (or query answer). 
Since DP naturally provides a convincing metric for quantifying privacy, i.e., privacy loss $\epsilon$, in recent years, some efforts have been put into pricing data based on data owners' privacy loss. 
Specifically, Ghosh et al. \cite{ghosh2011selling,ghosh2015selling} initiated the study of trading statistics over private data at auctions, perceiving the privacy loss as a commodity. 
Then, following \cite{li2012pricing}, Li et al. \cite{li2013theory,li2014theory} proposed an arbitrage-free pricing framework for noisy query answers perturbed by DP.
Similarly, considering machine learning models as query answers, Chen et al. \cite{chen2019towards} and Liu et al. \cite{liu2021dealer} proposed formal frameworks for pricing noisy versions of those models.
All the above DP based works assumed that data owners have to tolerate uniform and/or infinite privacy losses. 
However, Nget et al. \cite{nget2017balance} found that data owners have many diverse privacy preferences, and most of them value privacy more than money. 
Hence, in this paper, to achieve limited and personalized privacy losses, we allow each data owner to set a personalized upper bound of her own privacy loss and propose an arbitrage-free framework for trading noisy query answers with personalized privacy losses. 

\section{Conclusion}
\label{sec:conclusion}
We proposed a framework for continuously trading personal data where each data owner's privacy loss under personalized differential privacy is bounded and the arbitrage freeness constraint can be reasonably relaxed under bounded utilities by partial arbitrage freeness. 
To avoid arbitrage, we proposed theorems presenting sufficient conditions for guaranteeing arbitrage freeness and partial arbitrage freeness.
Based on those theorems, we designed two arbitrage-free trading protocols and two partially arbitrage-free protocols by instantiating the key modules in our framework. 
Finally, our experimental results showed that, the proposed PersonalizedTrading and PersonalizedTradingPlus protocols outperformed the UniformTrading protocol in terms of making full use of privacy loss bounds for data owners and could thus prevent arbitrage attacks.



%
%

\bibliographystyle{spbasic}      
\bibliography{ref}   


\appendix

\section{Missing Proofs}
\label{sec:proofs}

\begin{proofnoof}[Theorem \ref{theorem:arbitrage free}]
First, according to the first and second properties, we can construct a function $g=G(\theta)=\eta([f_1(\theta),...,f_n(\theta)])=\frac{v}{\lambda^2(\mathbf{q})}$. We drop the subscripts $\mathcal{M}$ from $U_{\mathcal{M}}$ and $\Pi_{\mathcal{M}}$ to simplify notation. Then, we construct a function $h(x)$ ($x\geq0$), which we will later prove subadditive on $x$, i.e., $\forall x_1, x_2 \geq 0$, $h(x_1)+h(x_2)\geq h(x_1+x_2)$.

Let $h(x)=
\begin{cases}
G^{-1}(\frac{1}{x}), & x>0\\
0, & x=0
\end{cases}$. 

Because of the fourth property in the theorem, we can derive that $\forall g>0, {(G^{-1})}^{\prime}(g)<0$. Then, because of the fifth property in the theorem, we have $h^{\prime \prime}(x) \leq 0$ for $x>0$:

\begin{align*}
&\forall \theta > 0, v \cdot \frac{\partial^2 v}{\partial \theta^2}\leq 2(\frac{\partial v}{\partial \theta})^2\\ 
\Rightarrow  &\forall \theta > 0, g\cdot G^{\prime \prime}(\theta)-2\cdot (G^{\prime}(\theta))^2\leq 0\\
\Rightarrow & \forall g>0, g \cdot [-\frac{{G^{-1}}^{\prime \prime}(g)}{({G^{-1}}^{\prime}(g))^3}] - 2\cdot(\frac{1}{{G^{-1}}^{\prime}(g)})^2 \leq 0 \\
\Rightarrow  &\forall g>0, g \cdot {G^{-1}}^{\prime \prime}(g) + 2\cdot {G^{-1}}^{\prime}(g) \leq 0\\
\Rightarrow & \forall x>0, \frac{1}{x} \cdot {G^{-1}}^{\prime \prime}(\frac{1}{x}) + 2\cdot {G^{-1}}^{\prime}(\frac{1}{x}) \leq 0 \\
\Rightarrow & \forall x>0, \frac{1}{x^4} \cdot {G^{-1}}^{\prime \prime}(\frac{1}{x}) + \frac{2}{x^3} \cdot {G^{-1}}^{\prime}(\frac{1}{x}) \leq 0\\
\Rightarrow & \forall x>0, h^{\prime \prime}(x)={G^{-1}}^{\prime \prime}(\frac{1}{x})\cdot [(\frac{1}{x})^{\prime}]^2 + {G^{-1}}^{\prime}(\frac{1}{x})\cdot (\frac{1}{x})^{\prime\prime}\leq 0
\end{align*}

Then, because of the third property in the theorem, $h(x)$ is right-continuous for $x=0$:
\begin{align*}
&\lim_{\theta->0^+}U(\mathbf{q}, [\epsilon_1,...,\epsilon_n])=+\infty \Rightarrow \lim_{\theta->0^+}G(\theta)=+\infty\\
\Rightarrow &\lim_{x->0^+}h(x)=\lim_{x->0^+}G^{-1}(\frac{1}{x})=\lim_{g->+\infty}G^{-1}(g)=0 \\
\Rightarrow &\lim_{x->0^+}h(x)=h(0)
\end{align*}

Then, according to the Lagrange mean value theorem, it can be implied that $h(x)$ is a subadditive function:

\begin{align*}
&\forall x_1,x_2 \in [0,+\infty), x_1\leq x_2,\\
&h(x_1)+h(x_2)-h(x_1+x_2)\\
=&[h(x_1)-h(0)]-[h(x_1+x_2)-h(x_2)]\\
=&x_1\cdot h^{\prime}(\xi_1)-x_1\cdot h^{\prime}(\xi_2)=x_1\cdot(\xi_1-\xi_2)\cdot h^{\prime\prime}(\xi_3)\geq 0
\end{align*}

where $\xi_1 \in (0,x_1)$, $\xi_2 \in (x_2,x_1+x_2)$, and $\xi_3 \in (\xi_1,\xi_2)$. 

Then, for any $\{\mathbf{q}_1, v_1\},...,\{\mathbf{q}_m, v_m\}$ and any $a_1,...,a_m$ such that $\sum_{j=1}^m a_j \mathbf{q}_j=\mathbf{q}$ and $\sum_{j=1}^m a_j^2v_j\leq v$, we have $h(\frac{\lambda^2(\mathbf{q})}{v})
\leq \sum_{j=1}^m h(\frac{\lambda^2(\mathbf{q}_j)}{v_j})$:

\begin{align*}
&h(\frac{\lambda^2(\mathbf{q})}{v})
\leq h(\frac{\lambda^2(\sum_{j=1}^m a_j \mathbf{q}_j)}{\sum_{j=1}^m a_j^2v_j})
\leq h(\frac{(\sum_{j=1}^m  \lambda(a_j \mathbf{q}_j))^2}{\sum_{j=1}^m a_j^2v_j}) \\
=& h(\frac{(\sum_{j=1}^m  |a_j|\lambda(\mathbf{q}_j))^2}{\sum_{j=1}^m a_j^2v_j})
\leq h(\frac{\sum_{j=1}^m a_j^2v_j)\cdot (\sum_{j=1}^m \frac{\lambda^2(\mathbf{q}_j)}{v_j})}{(\sum_{j=1}^m a_j^2v_j})\\
=& h(\sum_{j=1}^m \frac{\lambda^2(\mathbf{q}_j)}{v_j})\leq \sum_{j=1}^m h(\frac{\lambda^2(\mathbf{q}_j)}{v_j})
\end{align*}

Finally, we prove that $\pi=\Pi(v)$ is arbitrage free:

\begin{align*}
&\Pi(\mathbf{q}, v)
=\boldsymbol{\mu}((U^{\mathbf{q}}_{\mathcal{M}})^{-1}(v))
=\boldsymbol{\mu}(\eta^{-1}(\frac{v}{\lambda^2(\mathbf{q})}))\\
=&\boldsymbol{\mu}([f_1(G^{-1}(\frac{v}{\lambda^2(\mathbf{q})})),...,f_n(G^{-1}(\frac{v}{\lambda^2(\mathbf{q})}))])\\
=&\sum_i \mu_i(f_i(h(\frac{\lambda^2(\mathbf{q})}{v})))\\
\leq &\sum_i \mu_i(f_i(\sum_{j=1}^m h(\frac{\lambda^2(\mathbf{q}_j)}{v_j}))) \text{ (because $f_i$ and $\mu_i$ are increasing)}\\
\leq & \sum_{j=1}^m \sum_i \mu_i(f_i( h(\frac{\lambda^2(\mathbf{q}_j)}{v_j}))) \text{ (because $f_i$ and $\mu_i$ are subadditive)} \\
= &\sum_{j=1}^m \Pi(\mathbf{q}_j, v_j)
\end{align*}

\end{proofnoof}

\begin{proofnoof}[Theorem \ref{thm:equiv_pattern}]
Let $\rho_a, \rho_b$ be two exchangeable elements of pattern $\rho$. If we exchange $\rho_a, \rho_b$ to derive an equivalent pattern $\rho'$ of $\rho$, we have $\rho_a'=\rho_b$ and $\rho'_b=\rho_a$. May wish to assume $a=1$ and $b=2$ for simplicity. Then, because $\mu_a(\epsilon)=\mu_b(\epsilon)$ for all $\epsilon$, we have:
\begin{align*}
&\pi=\Pi^{\rho}_{Sam}(\boldsymbol{q},v)
=\boldsymbol{\mu}(\rho\cdot \theta_{max})\\
=&\mu_a(\rho_a\cdot\theta_{max})+\mu_b(\rho_b\cdot\theta_{max})+\sum_{i=3}^{n} \mu_i(\rho_i\cdot\theta_{max})\\
=&\mu_b(\rho'_b\cdot\theta_{max})+\mu_a(\rho'_a\cdot\theta_{max})+\sum_{i=3}^{n} \mu_i(\rho'_i\cdot\theta_{max})\\
=&\boldsymbol{\mu}(\rho'\cdot \theta_{max})=\Pi^{\rho'}_{Sam}(\boldsymbol{q},v)
\end{align*}
Therefore, $\Pi^{\rho}_{Sam}, \Pi^{\rho'}_{Sam}$ are equivalent.

Similarly, we can prove that if we exchange two exchangeable elements of $\rho'$ to derive an equivalent pattern $\rho''$ of $\rho$ and $\rho'$, the pricing functions $\Pi^{\rho}_{Sam}, \Pi^{\rho'}_{Sam}, \Pi^{\rho''}_{Sam}$ are equivalent.
Therefore, for any equivalent pattern $\rho'$ of $\rho$, the pricing functions $\Pi^{\rho}_{Sam}, \Pi^{\rho'}_{Sam}$ are equivalent.
\end{proofnoof} 

\begin{proofnoof}[Theorem \ref{thm:partial_ab}]
\label{proof:partial_ab}
First, according to the first and second properties, we can construct a function $g=G(\theta)=\eta([f_1(\theta),...,f_n(\theta)])=\frac{v}{\lambda^2(\mathbf{q})}$. We drop the subscripts $\mathcal{M}$ from $U_{\mathcal{M}}$ and $\Pi_{\mathcal{M}}$ to simplify notation. 
Then, let $h(x)=G^{-1}(\frac{1}{x}), x \in [\frac{1}{G(\theta^{L})}, \frac{1}{G(\theta^{U})}]$. 


Because of the fourth property in the theorem, we can derive that $\forall g > 0, {(G^{-1})}^{\prime}(g)<0$. Then, because of the fifth property in the theorem, we have $h^{\prime \prime}(x) \leq 0$ for $x \in [\frac{1}{G(\theta^{L})}, \frac{1}{G(\theta^{U})}]$:

\begin{align*}
&\forall \theta \in [\theta^{L}, \theta^{U}], v \cdot \frac{\partial^2 v}{\partial \theta^2}\leq 2(\frac{\partial v}{\partial \theta})^2 \\
\Rightarrow & \theta \in [\theta^{L}, \theta^{U}], g\cdot G^{\prime \prime}(\theta)-2\cdot (G^{\prime}(\theta))^2\leq 0\\
\Rightarrow & \forall g \in [G(\theta^{U}), G(\theta^{L})], g \cdot [-\frac{{G^{-1}}^{\prime \prime}(g)}{({G^{-1}}^{\prime}(g))^3}] - 2\cdot(\frac{1}{{G^{-1}}^{\prime}(g)})^2 \leq 0\\
\Rightarrow & \forall g \in [G(\theta^{U}), G(\theta^{L})], g \cdot {G^{-1}}^{\prime \prime}(g) + 2\cdot {G^{-1}}^{\prime}(g) \leq 0\\
\Rightarrow & \forall x \in [\frac{1}{G(\theta^{L})}, \frac{1}{G(\theta^{U})}], \frac{1}{x} \cdot {G^{-1}}^{\prime \prime}(\frac{1}{x}) + 2\cdot {G^{-1}}^{\prime}(\frac{1}{x}) \leq 0\\
\Rightarrow & \forall x \in [\frac{1}{G(\theta^{L})}, \frac{1}{G(\theta^{U})}], \frac{1}{x^4} \cdot {G^{-1}}^{\prime \prime}(\frac{1}{x}) + \frac{2}{x^3} \cdot {G^{-1}}^{\prime}(\frac{1}{x}) \leq 0\\
\Rightarrow & \forall x \in [\frac{1}{G(\theta^{L})}, \frac{1}{G(\theta^{U})}], \\
& h^{\prime \prime}(x)={G^{-1}}^{\prime \prime}(\frac{1}{x})\cdot [(\frac{1}{x})^{\prime}]^2 + {G^{-1}}^{\prime}(\frac{1}{x})\cdot (\frac{1}{x})^{\prime\prime}\leq 0
\end{align*}

Then, let $F(\theta)=\frac{1}{G(\theta)}$. Because of the third property, we have $F(\theta) + F(\theta^L) - F(\theta + \theta^L) \leq 0$ for $\theta \geq \theta^L, \theta \leq \theta^U - \theta^L$. Because of the fifth property, we have $F''(\theta) \geq 0$ for $\theta \in [\theta^L, \theta^U]$. Therefore, according to the Lagrange mean value theorem, we can conclude the following:
\begin{align*}
    & \forall \theta_1, \theta_2 \geq \theta^L, \theta_1 \leq \theta_2, \theta_1 + \theta_2 \leq \theta^U\\
    & F(\theta_1) + F(\theta_2) -F(\theta_1+\theta_2) \\
    = & (F(\theta_1) - F(\theta^L)) - (F(\theta_1+\theta_2) - F(\theta_2 + \theta^L))\\
    &+ (F(\theta_2) + F(\theta^L) - F(\theta_2 + \theta^L)) \\
    = & (\theta_1 - \theta^L)(F'(\xi_a) - F'(\xi_b)) + (F(\theta_2) \\
    &+ F(\theta^L) - F(\theta_2 + \theta^L))\\
    = & (\theta_1 - \theta^L)(\xi_a-\xi_b)F''(\xi_c) \\
    &+ (F(\theta_2) + F(\theta^L) - F(\theta_2 + \theta^L)) \leq 0
\end{align*}
where $\xi_a \in (\theta^{L},\theta_1)$, $\xi_b \in (\theta_2+\theta^{L},\theta_1+\theta_2)$, and $\xi_c \in (\xi_a,\xi_b)$. 

Then, because of the third property, we have $h(x)+h(F(\theta^{L}))-h(x+ F(\theta^{L})) \geq 0$:
\begin{align*}
&U(\boldsymbol{q},[f_1(2\theta^{L}),...,f_n(2\theta^{L})])\leq U(\boldsymbol{q},[f_1(\theta^{L}),...,f_n(\theta^{L})])/2 \\
\Rightarrow & G(2\theta^{L})\leq G(\theta^{L})/2 \\
\Rightarrow & 2\theta^{L} \geq G^{-1}(\frac{G(\theta^{L})}{2})\\
\Rightarrow  &2 G^{-1}(G(\theta^{L})) - G^{-1}(\frac{G(\theta^{L})}{2}) \geq 0 \\
\Rightarrow  &2h(\frac{1}{G(\theta^{L})}) - h(\frac{2}{G(\theta^{L})}) \geq 0\\
& \text{Let } k(x) = h(x) + h(F(\theta^{L})) - h(x + F(\theta^{L}))\\
& (x \geq F(\theta^L), x \leq F(\theta^U-\theta^L) ), \\
& k^{\prime}(x)=h^{\prime}(x) - h^{\prime}(x+F(\theta^{L})) \geq 0 \\
\Rightarrow & k(x) \geq k(F(\theta^{L})) \geq 0 \\
\Rightarrow & h(x)+h(F(\theta^{L}))-h(x+F(\theta^{L})) \geq 0\\
&(\forall x \geq F(\theta^L), x \leq F(\theta^U-\theta^L))
\end{align*}

Additionally, according to the Lagrange mean value theorem, it can be implied that:
\begin{align*}
&\forall x_1,x_2 \geq F(\theta^{L}), x_1\leq x_2 \leq F(\theta^U-\theta^L), x_1 + x_2 \leq F(\theta^{U})\\
&h(x_1)+h(x_2)-h(x_1+x_2)\\
=&(x_1-F(\theta^L))\cdot(\xi_1-\xi_2)\cdot h^{\prime\prime}(\xi_3) \\
& +  [h(x_2)+h(F(\theta^L))-h(x_2+F(\theta^{L}))]\geq 0
\end{align*}

where $\xi_1 \in (F(\theta^{L}),x_1)$, $\xi_2 \in (x_2+F(\theta^{L}),x_1+x_2)$, and $\xi_3 \in (\xi_1,\xi_2)$. 

Let $\psi(\boldsymbol{q}, \theta)=U_\mathcal{M}(\boldsymbol{q},[f_1(\theta),..., f_n(\theta)])$. Then, for any $(\boldsymbol{q},v)$ with $v\in [\psi(\boldsymbol{q},\theta^{U}), \psi(\boldsymbol{q},\theta^L)]$, for any $(\mathbf{q}_1, v_1),...,(\mathbf{q}_m, v_m)$ with each $v_j = \psi(\boldsymbol{q}_j,\theta_j), \theta_j \geq \theta^L$ and $\sum_{j=1}^{m} \theta_j \leq \theta^U$, and for any $a_1,...,a_m$ such that $\sum_{j=1}^m a_j \mathbf{q}_j=\mathbf{q}$ and $\sum_{j=1}^m a_j^2v_j\leq v$, we have $h(\frac{\lambda^2(\mathbf{q})}{v})
\leq \sum_{j=1}^m h(\frac{\lambda^2(\mathbf{q}_j)}{v_j})$:

\begin{align*}
&h(\frac{\lambda^2(\mathbf{q})}{v})
\leq h(\frac{\lambda^2(\sum_{j=1}^m a_j \mathbf{q}_j)}{\sum_{j=1}^m a_j^2v_j})
\leq h(\frac{(\sum_{j=1}^m  \lambda(a_j \mathbf{q}_j))^2}{\sum_{j=1}^m a_j^2v_j}) \\
= & h(\frac{(\sum_{j=1}^m  |a_j|\lambda(\mathbf{q}_j))^2}{\sum_{j=1}^m a_j^2v_j})\leq h(\frac{(\sum_{j=1}^m a_j^2v_j)\cdot (\sum_{j=1}^m \frac{\lambda^2(\mathbf{q}_j)}{v_j})}{\sum_{j=1}^m a_j^2v_j})\\
= & h(\sum_{j=1}^m \frac{\lambda^2(\mathbf{q}_j)}{v_j}) = h(\sum_{j=1}^{m} F(\theta_j))
\leq \sum_{j=1}^m h(F(\theta_j)) = \sum_{j=1}^m h(\frac{\lambda^2(\mathbf{q}_j)}{v_j})
\end{align*}

Finally, we prove $\pi=\Pi(v)$ partially arbitrage free for $v\in[\psi(\boldsymbol{q},\theta^{U}), \psi(\boldsymbol{q},\theta^L)]$:

\begin{align*}
&\Pi(\mathbf{q}, v)=\boldsymbol{\mu}((U^{\mathbf{q}}_{\mathcal{M}})^{-1}(v))=\boldsymbol{\mu}(\eta^{-1}(\frac{v}{\lambda^2(\mathbf{q})}))\\
=&\boldsymbol{\mu}([f_1(G^{-1}(\frac{v}{\lambda^2(\mathbf{q})})),...,f_n(G^{-1}(\frac{v}{\lambda^2(\mathbf{q})}))])\\
=&\sum_i \mu_i(f_i(h(\frac{\lambda^2(\mathbf{q})}{v})))\leq \sum_i \mu_i(f_i(\sum_{j=1}^m h(\frac{\lambda^2(\mathbf{q}_j)}{v_j}))) \\
\leq &\sum_{j=1}^m \sum_i \mu_i(f_i( \sum_{j=1}^m h(\frac{\lambda^2(\mathbf{q}_j)}{v_j})))= \sum_{j=1}^m \Pi(\mathbf{q}_j, v_j)
\end{align*}
\end{proofnoof}

\begin{proofnoof}[Theorem \ref{thm:partial_ab_sup}]
First, according to the first and second properties, we can construct a function $g=G(\theta)=\eta([f_1(\theta),...,f_n(\theta)])=\frac{v}{\lambda^2(\mathbf{q})}$. We drop the subscripts $\mathcal{M}$ from $U_{\mathcal{M}}$ and $\Pi_{\mathcal{M}}$ to simplify notation. 
Then, for each data owner $u_i$, we construct a function $h_i(x)$. Let $h_i(x)=\mu_i(f_i(G^{-1}(\frac{1}{x})))$ with $x \in [\frac{1}{G(\theta^{L})}, \frac{1}{G(\theta^{U})}]$, $\forall i$. 

Because of the first property, we can derive that $\forall g > 0, {(G^{-1})}^{\prime}(g)<0$. Then, because of the third property in the theorem, we have $h_i^{\prime \prime}(x) \leq 0$ for $x \in [\frac{1}{G(\theta^{L})}, \frac{1}{G(\theta^{U})}], \forall i$ if $\mu_i$ is superadditive:

\begin{align*}
&\forall \theta \in [\theta^{L}, \theta^{U}], v \cdot \frac{\partial^2 v}{\partial c_i^2}\leq 2(\frac{\partial v}{\partial c_i})^2 \\
\Rightarrow &\forall \theta \in [\theta^{L}, \theta^{U}], g \cdot \frac{\partial^2 g}{\partial c_i^2}\leq 2(\frac{\partial g}{\partial c_i})^2\\
\Rightarrow &\forall \theta \in [\theta^{L}, \theta^{U}], \\
& g \cdot [G'(f^{-1}_i(\mu_i^{-1}(c_i)))\cdot ({\mu_i^{-1}}'(c_i))^2\cdot {f^{-1}_i}''(\mu_i^{-1}(c_i))\\
&+ ({f^{-1}_i}'(\mu_i^{-1}(c_i)))^2 \cdot ({\mu_i^{-1}}'(c_i))^2 \cdot G''(f^{-1}_i(\mu_i^{-1}(c_i)))\\
&+{f^{-1}_i}'(\mu_i^{-1}(c_i))\cdot G'(f^{-1}_i(\mu_i^{-1}(c_i)))\cdot {\mu_i^{-1}}''(c_i)]\\
&\leq 2({f^{-1}_i}'(\mu_i^{-1}(c_i))\cdot G'(f^{-1}_i(\mu_i^{-1}(c_i)))\cdot {\mu_i^{-1}}'(c_i))^2\\
\Rightarrow &\forall \theta\in [\theta^{L}, \theta^{U}], \\
& g \cdot [G'(\theta)\cdot ({\mu_i^{-1}}'(c_i))^2\cdot {f^{-1}_i}''(\epsilon_i)\\
& + ({f^{-1}_i}'(\epsilon_i))^2 \cdot ({\mu_i^{-1}}'(c_i))^2 \cdot G''(\theta)\\
&+{f^{-1}_i}'(\epsilon_i)\cdot G'(\theta)\cdot {\mu_i^{-1}}''(c_i)]\\
&\leq 2({f^{-1}_i}'(\epsilon_i)\cdot G'(\theta)\cdot {\mu_i^{-1}}'(c_i))^2\\
\Rightarrow &\forall i, \forall \theta \in [\theta^{L}, \theta^{U}], \\
& g \cdot [({G^{-1}}'(g))^{-1}\cdot ({\mu_i}'(\epsilon_i))^{-2}\cdot \frac{-{f_i}''(\theta)}{({f_i}'(\theta))^3}\\
&+ ({f_i}'(\theta))^{-2} \cdot ({\mu_i}'(\epsilon_i))^{-2} \cdot \frac{-{G^{-1}}''(g)}{({G^{-1}}'(g))^3}\\
&+({f_i}'(\theta))^{-1}\cdot ({G^{-1}}'(g))^{-1}\cdot \frac{-{\mu_i}''(\epsilon_i)}{({\mu_i}'(\epsilon_i))^3}]\\
&\leq 2({f_i}'(\theta)\cdot {G^{-1}}'(g)\cdot {\mu_i}'(\epsilon_i))^{-2}\\
\Rightarrow &\forall i, \forall \theta \in [\theta^{L}, \theta^{U}], ({G^{-1}}'(g))^{2}\cdot {\mu_i}'(\epsilon_i)\cdot {f_i}''(\theta)\\
&+ {f_i}'(\theta)\cdot {\mu_i}'(\epsilon_i) \cdot {G^{-1}}''(g)\\
&+({f_i}'(\theta))^{2}\cdot ({G^{-1}}'(g))^{2}\cdot {\mu_i}''(\epsilon_i) \\
&+ 2({f_i}'(\theta)\cdot {G^{-1}}'(g)\cdot {\mu_i}'(\epsilon_i)) /g\leq 0\\
\Rightarrow &\forall g \in [G(\theta^{U}), G(\theta^{L})],\\
& g^4 \cdot [({G^{-1}}'(g))^{2}\cdot {\mu_i}'(f_i(G^{-1}(g)))\cdot {f_i}''(G^{-1}(g))\\
&+ {f_i}'(G^{-1}(g))\cdot {\mu_i}'(f_i(G^{-1}(g))) \cdot {G^{-1}}''(g) \\
&+({f_i}'(G^{-1}(g)))^{2}\cdot ({G^{-1}}'(g))^{2}\cdot {\mu_i}''(f_i(G^{-1}(g)))\\
& + 2({f_i}'(\theta)\cdot {G^{-1}}'(g)\cdot {\mu_i}'(f_i(G^{-1}(g)))) /g]\leq 0\\
\Rightarrow &\forall x \in [\frac{1}{G(\theta^{L})}, \frac{1}{G(\theta^{U})}],\\
h_i''(x)& =x^{-4} \cdot [({G^{-1}}'(\frac{1}{x}))^{2}\cdot {\mu_i}'(f_i(G^{-1}(\frac{1}{x})))\cdot {f_i}''(G^{-1}(\frac{1}{x}))\\
&+ {f_i}'(G^{-1}(\frac{1}{x}))\cdot {\mu_i}'(f_i(G^{-1}(\frac{1}{x}))) \cdot {G^{-1}}''(\frac{1}{x})\\
&+({f_i}'(G^{-1}(\frac{1}{x})))^{2}\cdot ({G^{-1}}'(\frac{1}{x}))^{2}\cdot {\mu_i}''(f_i(G^{-1}(\frac{1}{x})))\\
& + 2({f_i}'(\theta)\cdot {G^{-1}}'(\frac{1}{x})\cdot {\mu_i}'(f_i(G^{-1}(\frac{1}{x})))) \cdot x]\leq 0
\end{align*}

Then, let $F(\theta)=\frac{1}{G(\theta)}$. Because of the first property, according to Proof \ref{proof:partial_ab}, we can conclude the following:
\begin{align*}
    & \forall \theta_1, \theta_2 \geq \theta^L, \theta_1 \leq \theta_2, \theta_1 + \theta_2 \leq \theta^U, \\
    & F(\theta_1) + F(\theta_2) -F(\theta_1+\theta_2) \leq 0
\end{align*}

Then, because of the second property, $\forall i$, if $\mu_i$ is superadditive, we have:
\begin{align*}
& U_{\mathcal{M}}(\boldsymbol{q},[f_1(\theta^{A}_i),...,f_n(\theta^{A}_i)])\leq \frac{U_{\mathcal{M}}(\boldsymbol{q},[f_1(\theta^{L}),...,f_n(\theta^{L})])}{2} \\
\Rightarrow & G(\theta^{A}_i)\leq \frac{G(\theta^{L})}{2}\\
\Rightarrow & G({f_i}^{-1}({\mu_i}^{-1}(2\mu_i(f_i(\theta^L))))) \leq\frac{G(\theta^{L})}{2}\\
\Rightarrow & 2\mu_i(f_i(\theta^L)) \geq \mu_i(f_i(G^{-1}(\frac{G(\theta^{L})}{2})))\\
\Rightarrow & 2\mu_i(f_i(G^{-1}(G(\theta^{L})))) - \mu_i(f_i(G^{-1}(\frac{G(\theta^{L})}{2}))) \geq 0 \\
\Rightarrow & 2h_i(\frac{1}{G(\theta^{L})}) - h_i(\frac{2}{G(\theta^{L})}) \geq 0\\
& \text{Let } k_i(x) = h_i(x) + h_i(F(\theta^{L})) - h_i(x + F(\theta^{L}))\\
& (x \geq F(\theta^L), x\leq F(\theta^U-\theta^L)), \\
& k_i^{\prime}(x)=h_i^{\prime}(x) - h_i^{\prime}(x+F(\theta^{L})) \geq 0 \\
\Rightarrow & k_i(x) \geq k_i(F(\theta^{L})) \geq 0 \\
\Rightarrow & h_i(x)+h_i(F(\theta^{L}))-h_i(x+F(\theta^{L})) \geq 0,\\
& (\forall x \geq F(\theta^L), x\leq F(\theta^U-\theta^L))
\end{align*}

According to the Lagrange mean value theorem, the following property holds for each $h_i$ if $\mu_i$ is superadditive:
\begin{align*}
&\forall x_1,x_2 \geq F(\theta^{L}), x_1\leq x_2 \leq F(\theta^U-\theta^L), x_1 + x_2 \leq F(\theta^{U})\\
&h_i(x_1)+h_i(x_2)-h_i(x_1+x_2)\\
=& (x_1-F(\theta^L))\cdot(\xi_1-\xi_2)\cdot h_i^{\prime\prime}(\xi_3) \\
& +  [h_i(x_2)+h_i(F(\theta^L))-h_i(x_2+F(\theta^{L}))]\geq 0
\end{align*}
where $\xi_1 \in (F(\theta^{L}),x_1)$, $\xi_2 \in (x_2+F(\theta^{L}),x_1+x_2)$, and $\xi_3 \in (\xi_1,\xi_2)$.


Let $\psi(\boldsymbol{q}, \theta)=U_\mathcal{M}(\boldsymbol{q},[f_1(\theta),..., f_n(\theta)])$. Then, for any $(\boldsymbol{q},v)$ with $v\in [\psi(\boldsymbol{q},\theta^{U}), \psi(\boldsymbol{q},\theta^L)]$, for any $(\mathbf{q}_1, v_1),...,(\mathbf{q}_m, v_m)$ with each $v_j = \psi(\boldsymbol{q}_j,\theta_j), \theta_j \geq \theta^L$ and $\sum_{j=1}^{m} \theta_j \leq \theta^U$, and for any $a_1,...,a_m$ such that $\sum_{j=1}^m a_j \mathbf{q}_j=\mathbf{q}$ and $\sum_{j=1}^m a_j^2v_j\leq v$, we have $h_i(\frac{\lambda^2(\mathbf{q})}{v})
\leq \sum_{j=1}^m h_i(\frac{\lambda^2(\mathbf{q}_j)}{v_j}), \forall i$, if $\mu_i$ is superadditive:
\begin{align*}
& h_i(\frac{\lambda^2(\mathbf{q})}{v}) \leq h_i(\frac{\lambda^2(\sum_{j=1}^m a_j \mathbf{q}_j)}{\sum_{j=1}^m a_j^2v_j})\leq h_i(\frac{(\sum_{j=1}^m  \lambda(a_j \mathbf{q}_j))^2}{\sum_{j=1}^m a_j^2v_j}) \\
= & h_i(\frac{(\sum_{j=1}^m  |a_j|\lambda(\mathbf{q}_j))^2}{\sum_{j=1}^m a_j^2v_j})\leq h_i(\frac{(\sum_{j=1}^m a_j^2v_j)\cdot (\sum_{j=1}^m \frac{\lambda^2(\mathbf{q}_j)}{v_j})}{\sum_{j=1}^m a_j^2v_j})\\
= & h_i(\sum_{j=1}^m \frac{\lambda^2(\mathbf{q}_j)}{v_j})
= h_i(\sum_{j=1}^{m} F(\theta_j))
\leq \sum_{j=1}^m h_i(F(\theta_j))\\
= & \sum_{j=1}^m h_i(\frac{\lambda^2(\mathbf{q}_j)}{v_j}),
\end{align*}
which also holds if $\mu_i$ is subadditive according to Proof \ref{proof:partial_ab}.

Finally, we prove $\pi=\Pi(v)$ partially arbitrage free for $v\in[\psi(\boldsymbol{q},\theta^{U}), \psi(\boldsymbol{q},\theta^L)]$:

\begin{align*}
&\Pi(\mathbf{q}, v) =\boldsymbol{\mu}((U^{\mathbf{q}}_{\mathcal{M}})^{-1}(v))
=\boldsymbol{\mu}(\eta^{-1}(\frac{v}{\lambda^2(\mathbf{q})}))\\
=& \boldsymbol{\mu}([f_1(G^{-1}(\frac{v}{\lambda^2(\mathbf{q})})),...,f_n(G^{-1}(\frac{v}{\lambda^2(\mathbf{q})}))])\\
=&\sum_i \mu_i(f_i(G^{-1}(\frac{v}{\lambda^2(\mathbf{q})}))) = \sum_i h_i(\frac{\lambda^2(\mathbf{q})}{v})\\
\leq & \sum_i \sum_{j=1}^m h_i(\frac{\lambda^2(\mathbf{q}_j)}{v_j})\leq \sum_{j=1}^m \sum_i h_i(\frac{\lambda^2(\mathbf{q}_j)}{v_j})= \sum_{j=1}^m \Pi(\mathbf{q}_j, v_j)
\end{align*}
\end{proofnoof}

\end{document}